\documentclass[%
 reprint,
 amsmath,amssymb,
 aps,
 pra,
]{revtex4-2}

\usepackage[utf8]{inputenc}
\usepackage{graphicx}% Include figure files
\usepackage{dcolumn}% Align table columns on decimal point
\usepackage{bm}% bold math
\usepackage[dvipsnames]{xcolor} % for textcolor{color}{red}
\usepackage{siunitx}
\usepackage{comment}
\usepackage{color}

\usepackage{placeins}
\usepackage[english]{babel}
\usepackage{gensymb} % for degrees

\begin{document}
%\nolinenumbers

\title{Space-Charge Effects During Half-Integer Resonance Crossing in the CERN PSB}

\author{Tirsi Prebibaj}
  \email{tirsi.prebibaj@cern.ch}
\author{Fanouria Antoniou, Foteini Asvesta, Hannes Bartosik}
\affiliation{CERN, Geneva, Switzerland}
\author{Giuliano Franchetti}
\affiliation{
GSI Helmholtzzentrum f\"ur Schwerionenforschung GmbH,
64291 Darmstadt, Germany, \\
Goethe University Frankfurt, Max-von-Laue-Stra\ss e 7, 60438 Frankfurt am Main, Germany \\
Helmholtz Forschungsakademie Hessen für FAIR (HFHF), GSI Helmholtzzentrum für Schwerionenforschung, Campus Frankfurt, Frankfurt am Main, Germany.
}

%\date{\today}

\begin{abstract}

The survival of charged particles in synchrotrons requires avoiding setting the beam on machine resonances, the most dangerous of which are the integer and half-integer. Nevertheless, operationally, the transverse tunes may change dynamically, crossing these resonances, and resulting in unwanted beam quality degradation and beam loss. For high intensity beams the process of resonance crossing is even more critical as space charge, in addition to the incoherent effects, may generate also coherent effects on the beam envelope dynamics. The interplay of the speed of resonance crossing, the space charge incoherent tune spread, and half-integer resonance width are fundamental for the beam behavior. In this paper we study the crossing of the half-integer resonance $2Q_y=9$ in the CERN Proton Synchrotron Booster~(PSB) for a range of parameters using a coasting beam. We take advantage of the newly commissioned PSB to investigate the beam response experimentally. The experimental findings are also discussed with the support of simulations.

\end{abstract}

\maketitle

%\tableofcontents

\section{Introduction}\label{sec:introduction}

It is well known that the effect of magnetic error resonances on the beam dynamics in accelerators becomes significant when the machine tunes $Q_x,Q_y$ satisfy the resonance condition $n_x Q_x + n_y Q_y=N$~\cite{Guignard:1978jm}. Here $n_x,n_y$ are integers, and $|n_x|+|n_y|$ is the resonance order, $N$ is the harmonics of the non-linear components driving the resonance. From the theory of resonances~\cite{Guignard:1978jm}, it is found that the low-order machine resonances, such as integer and half-integer, are the most detrimental for the beam dynamics. Dipolar errors create closed orbit distortions, whose amplitude becomes very large when the tunes approach the corresponding integer. In this paper, we focus on the effects of gradient errors, which, for a pure linear machine may: 1) affect the transverse dynamics and optics: if $M_y$ is the vertical one-turn map of the unperturbed machine, which has $|\textrm{Tr}(M_y)| < 2$ for every $Q_y$ but for $2Q_y=N$, then the presence of gradient errors creates a region around $2Q_y=N$ where $|\textrm{Tr}(M_y)| > 2$. This is a stopband of unstable single-particle motion. In this sense, gradient errors do not excite a resonance, but create a single particle instability. At the same time, the gradient errors may create a beta-beat that diverges as $Q_y$ approaches the half-integer tune. Courant-Snyder discuss this effect in terms of a resonant process created by the betatron functions~\cite{COURANT19581}. 2) Skew gradient errors instead couple the horizontal and vertical planes. For tunes satisfying $Q_x \pm Q_y = N$ a resonance condition is met. However, this is not a resonance, in fact, the extension of the Courant-Snyder theory developed by Edward and Teng~\cite{Edwards-Teng:1973}  allows a full parameterization of the motion without invoking any resonant dynamics. For a beam having RMS emittances $\epsilon_x \gg \epsilon_y$, the main feature of the linear coupling is to create an emittance exchange when the tunes are set close to $Q_x-Q_y=N$~\cite{PhysRevAccelBeams.23.044003}. A stopband of unstable motion can also appear for a $4\times 4$ one turn map if the modulus of its eigenvalue is larger than one~\cite{Floquet1884}. 

Therefore, for an ideal linear lattice, the effect of additional gradient errors is the creation of unstable linear motion or to linearly couple the two planes of motion. 

When considering a more realistic lattice in which also magnet non-linear components are present, the single particle dynamics further change. The non-resonant non-linear components from magnets create an amplitude-dependent detuning, which changes the nature of the dynamics. In this case, the Courant-Snyder or Edward-Teng theories apply only at first order, and one should discuss the particle motion in the domain of the non-linear dynamics. The main methods to analyze this situation are the perturbative approaches~\cite{Guignard:1978jm} or the normal form theory~\cite{Bazzani:1994ks}. Applying the analysis, it turns out that for this type of lattices, the overall effect of a 1D resonance is the creation of islands in phase space.  Mastering the location of the islands and their size has become a standard in accelerator physics: novel schemes of beam extraction are based on precise control of islands and separatrixes, and have been deployed to improve the beam extraction~\cite{Giovannozzi:2025zmf, PhysRevResearch.6.L042018, PhysRevLett.88.104801,   slowextractionworkshops}. The key point in this scenario is that the location of the island's fixed point is entirely determined by the resonance condition on the single particle dynamics, and so the amplitude-dependent detuning controls the fixed points in phase space. 

For high-intensity beams, the dynamics is more complex, as the Coulomb forces create a smooth, long-range electric field (space charge), which couples the dynamics of a single particle to the dynamics of the beam as a whole (we neglect here intra-beam scattering). Consequently, the evolution of the full beam has to be treated using the Vlasov approach \cite{vlasov}. In beam physics, the conundrum arises when the coherent dynamics originating from the Vlasov equation is affected by the machine gradient errors, which generate incoherent forces on all beam particles. 

If the beam distribution is stationary, then for a particle distribution encountered in a real machine (e.g.~Gaussian distribution) the space charge produces an amplitude-dependent detuning, which is also stationary, and can affect the dynamics as any other non-resonant non-linear element does. In contrast to the magnet elements, the space charge produces a larger tune-shift $\delta Q_{x}^{SC},\delta Q_{y}^{SC}$ at the beam center, whereas particles having larger oscillation amplitudes experience a smaller tune-shift, which altogether form the space charge tune-spread. Under the assumption that the beam distribution remains stationary, one can discuss the effect of the ``incoherent'' space charge on the resonant dynamics, and at which amplitudes particles would become resonant. This view is adopted in a conservative analysis of the beam dynamics: the machine tunes are set so that the incoherent space charge tune-spread does not overlap with any machine resonance, especially with integer and half-integer resonances. This view assumes that the coherent and collective beam behavior is negligible. 

The coherent beam behavior has been discussed first by Frank Sacherer in a scenario with an intense KV coasting beam responding to a gradient error~\cite{Sacherer:1968xh}. He has shown that for a small gradient error the effect of the resonance does not appear at the shift dictated by $\delta Q_{y}^{SC}$, but rather at a different tune-shift, a ``coherent'' tune-shift which relate to the incoherent with $\delta Q_{y}^{coh} = 3/4 \delta Q_{y}^{SC}$. This result shows that the intense beam responds as a whole to an incoherent resonance, resulting in a coherent beam response. In analogy to the case of the single particle dynamics, one can discuss the dynamics of the beam envelope, and find that also the envelope can be affected by instabilities. The gradient errors perturb linearly the beam envelope, and the envelope equation manifestly shows that the envelope becomes unstable when the coherent tunes meet a resonance condition. An extension of the envelope stability is generalized in \cite{PhysRevE.57.4713}, which experimental investigation has been explored in traps \cite{PhysRevSTAB.13.044201,Sheehy:IPAC2016-WEPOY048} to verify the properties of the coherent resonances. 

These coherent beam effects are based on the assumption that the beam distribution is uniform (KV distribution~\cite{reiser2008theory}), and so a perturbative approach to the Vlasov equation can be carried out to characterize the stability properties of the beam envelope. However, for a realistic non-KV beam, the spatial beam profile is non-uniform and the situation is more complex. The beam RMS equivalence found by Sacherer~\cite{Sacherer:322516} suggests that the beam may respond coherently to gradient errors, whereas the non-linear beam distribution generates a single-particle, amplitude-dependent tune that determines the resonant dynamics of individual particles located in the tails or in the beam core. The dominance of one effect over the other depends on the specific scenario, and the consequences are significant for practical beam operations.

The half-integer resonance for high-intensity beams is a subject of great interest, as it directly affects the performance of beam delivery. For example, at Fermilab National Lab~(FNAL), an extensive campaign of studies in the Fermilab Booster has investigated the half-integer resonance~\cite{Eldred_2024}. Proposals for its correction in the Booster are discussed in~\cite{valishev_hb2016}, while studies of characterization and optimization strategies are presented in \cite{rabusov,rabusov_nima}. At the Rutherford Appleton Laboratory~(RAL), several studies have also been performed to understand the effects of the half-integer resonance and models for beam loss have been proposed using a frozen model \cite{Warsop_hb2016}.

In the context of FFAs (Fixed Field Accelerators)~\cite{machida2012emma}, acceleration induces a dynamic shift in the machine tunes, resulting in a single-passage resonance crossing, opposite to the periodic crossing induced in a synchrotron bunch. This effect was studied in Ref.~\cite{Lee_2006}. We explore here a similar scenario, which can occur in standard  operation of synchrotrons: the beam is injected at a given working point and then shifted as rapidly as possible to another working point, crossing the half-integer resonance in the process. In this situation, the machine tunes cross a half-integer resonance, and the beam-resonance interaction becomes complex. An incoherent effect is expected due to the dynamic change in tune, which may cause particles to become trapped in two resonance islands. On the other hand, a coherent instability may also arise, altering the overall beam response.

In this paper, we report on an investigation carried out at the Proton Synchrotron Booster (PSB) at CERN, where a proof-of-principle setup was implemented to study the half-integer resonance crossing. In this experiment, a controlled gradient error was deliberately introduced to excite the $2Q_y =9$ resonance in an otherwise half-integer corrected lattice. This intentional perturbation created a controlled resonance crossing scenario, allowing us to explore the interplay between incoherent particle dynamics and coherent/collective beam behavior during the resonance crossing process using a coasting beam.

The PSB is a four-ring synchrotron serving as the first circular accelerator after Linac4 in the CERN proton injector chain for the LHC and other fixed-target facilities. With the completion of the LHC Injectors Upgrade~(LIU) project~\cite{liu_tdr_v1}, the achievable brightness in the PSB was roughly doubled, while maintaining excellent control over beam and machine parameters, including intensity, emittance, tune, chromaticity, and resonance correctors. This precise machine control enabled the systematic study of the half-integer resonance crossing, which is also of operational interest for exploring the PSB brightness limits.

Two resonance crossings are explored: ``crossing from above'', where the vertical tune~$Q_y$ starts above the half-integer resonance and is programmed to dynamically decrease in a fixed energy machine cycle, and ``crossing from below'', where~$Q_y$ starts below and is programmed to dynamically increase. The machine tune, as set by the quadrupole magnet currents, matches with the tune measured from the turn-by-turn beam centroid motion~\cite{Gasior:883298,Gasior:2019hxg} (center-of-mass tune). This gives us the confirmation that transverse impedance and image charges are negligible in this experimental campaign. 

All measurements are compared to 6D tracking simulations performed with \texttt{Xsuite}~\cite{Iadarola:2023fuk}. For all experiments, the simulated macroscopic beam parameters (intensity, emittances, momentum spread) were adjusted in the simulations to match the measured ones. The tracking is based on a detailed PSB lattice model~\cite{cern_acc_models}, including the physical aperture and relevant machine settings, such as tunes, chromaticity, and resonance excitation. Direct space charge was included via a Particle-In-Cell~(PIC) solver, while other collective effects (indirect space charge, impedance) were neglected. A more detailed description of the simulation setup is provided in Appendix~\ref{sec:experimental_setup}.

The paper is organized as follows. Section~\ref{sec:half-integer_resonance_control} describes the control and measurement of the half-integer resonance. The results for resonance crossing from above and from below are presented in Sec.~\ref{sec:half-integer_resonance_crossing_above} and Sec.~\ref{sec:half-integer_resonance_crossing_below}, respectively. Finally, Sec.~\ref{sec:conclusions_outlook} summarizes the findings and conclusions.

\section{Half-Integer Resonance Control}\label{sec:half-integer_resonance_control}

The study focuses on the vertical half-integer resonance $2Q_y = 9$, which occurs when the vertical tune is $Q_y = 4.5$. At this tune, the phase advance per cell is $101.25\degree$. Because the PSB lattice has a periodicity of 16, this resonance is not structural and can be driven by random quadrupole-like field errors.

The vertical half-integer resonance $2Q_y=9$ in the PSB has been extensively characterized in previous studies~\cite{PhysRevAccelBeams.19.124202,prebibaj:ipac2022-mopost058}. Two families of quadrupole correctors, QNO412 and QNO816, generate orthogonal driving terms which are used to compensate residual quadrupole-like errors. Within each family, the correctors are arranged with opposite polarities and symmetric locations, allowing them to be powered without changing the overall machine tune. 

The resonance compensation is performed experimentally~\cite{SantamariaGarcia:2019rrg} by scanning the currents of both families and identifying the setting that minimizes beam losses when crossing the resonance with a low-intensity beam.

\begin{figure}[!htb]
   \centering
   \includegraphics*[width=0.99\columnwidth]{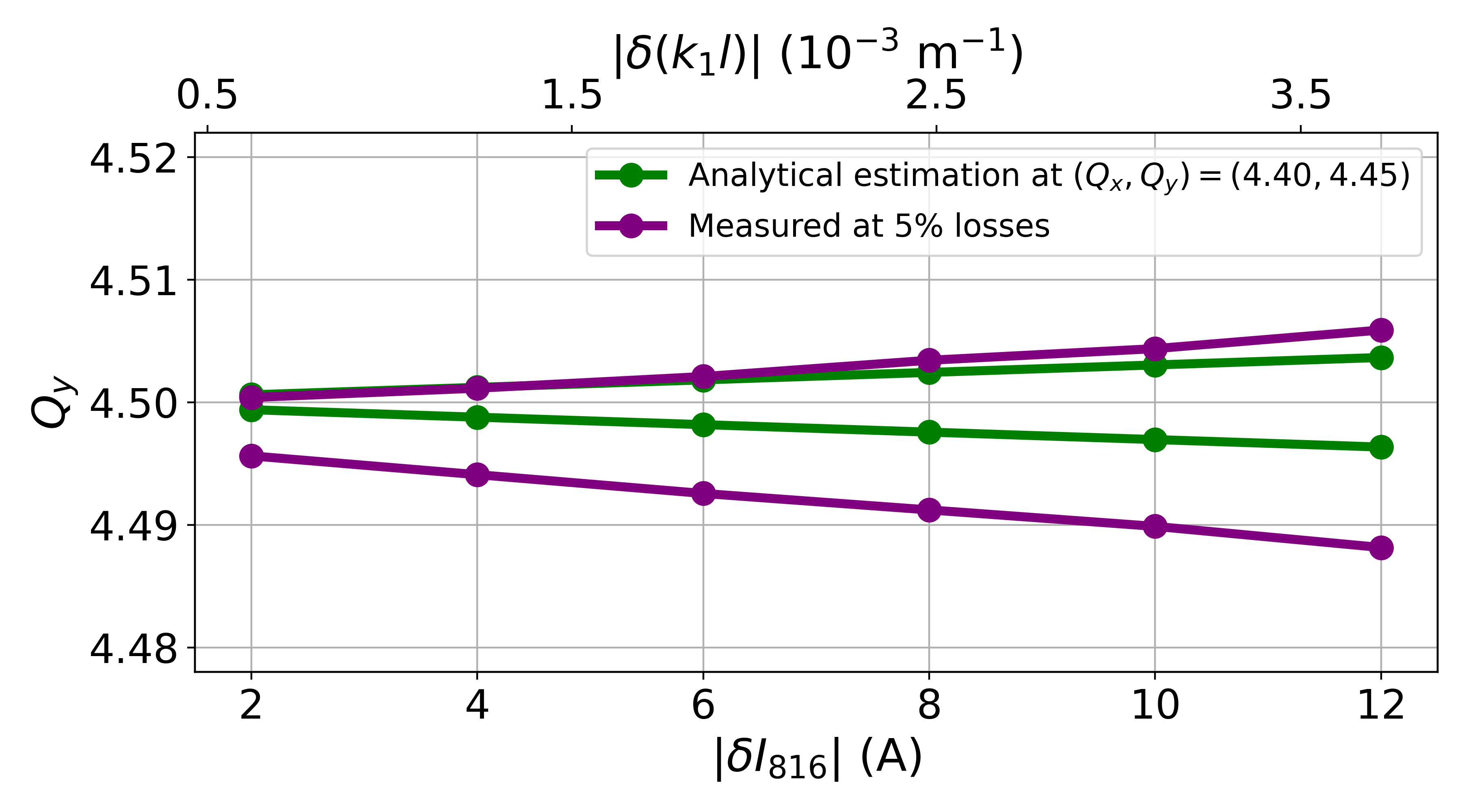}
   \caption{Experimentally determined half-integer stopband upper and lower boundaries (purple), and comparison with analytical estimations (green), for different resonance excitation levels. The excitation corresponds to the absolute current shift of the QNO816 quadrupole family with respect to its compensating value. A shift of $| \delta I_\textrm{816}| = 2$~A corresponds to $|\delta (k_\textrm{1} l)| = 0.62\times 10^{-3}$\,$\textrm{m}^{-1}$.}
   \label{fig:half-integer_resonance_stopband_widths}
\end{figure}

Once compensated, the resonance can be intentionally excited by varying the corrector strengths relative from their compensating values. The corresponding stopband width is then determined experimentally, as described in~\cite{prebibaj:ipac2022-mopost058} and shown in Fig.~\ref{fig:half-integer_resonance_stopband_widths}. For a given excitation, a low-intensity beam is used to cross the resonance from both directions while recording beam losses. The stopband boundaries are defined as the tune where 5\% losses occur, providing an experimental estimate of the stopband width. Although the actual width may be slightly smaller, this method gives a reasonable estimate for the present study. For the weakest excitation, the resonance stopband width is $\delta Q_{y,\mathrm{res}} \approx 0.004$.

\section{Half-Integer Crossing from Above the Resonance}\label{sec:half-integer_resonance_crossing_above}

\subsection{Crossing without external resonance excitation}\label{subsec:no_trapping}

\begin{figure}[!htb]
   \centering
   \includegraphics*[width=.99\columnwidth]{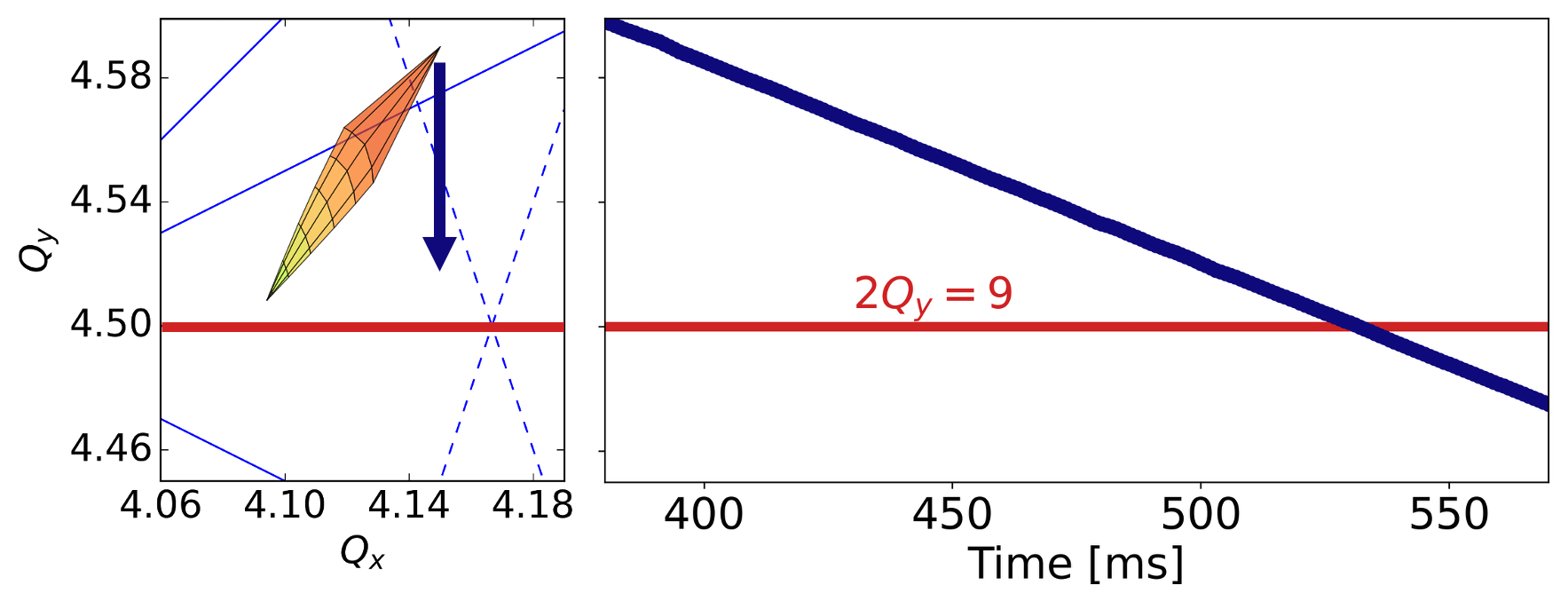}
   \caption{Left: working point evolution (blue arrow) and analytically estimated space charge tune spread (orange polygons)~\cite{PySCRDT}. Right: measured vertical tune as a function of time during the crossing of the half-integer resonance $2Q_y=9$. 1~ms at PSB flat-bottom energy corresponds to approximately 1000 beam revolutions.}
   \label{fig:tune_ramp}
\end{figure}

A first experimental measurement was performed with the half-integer resonance compensated, i.e.~without applying intentional resonance excitation after finding the best corrector settings for resonance compensation (as described in Sec.~\ref{sec:half-integer_resonance_control}). In this configuration, the half-integer resonance was crossed from above with a tune change per turn (crossing rate) of approximately $\delta Q_y / \text{turn} = -0.66 \times 10^{-6}$, as shown in Fig.~\ref{fig:tune_ramp}. A coasting beam with a maximum incoherent space charge tune shift of $\delta Q_y^{SC} \approx -0.09$ was used and the vertical chromaticity was experimentally compensated using sextupole correctors. The full parameters of the experimental setup are listed in Table~\ref{tab:beam_params_crossing_above} of Appendix~\ref{sec:experimental_setup}.

\begin{figure}[!htb]
   \centering
   \includegraphics*[width=.99\columnwidth]{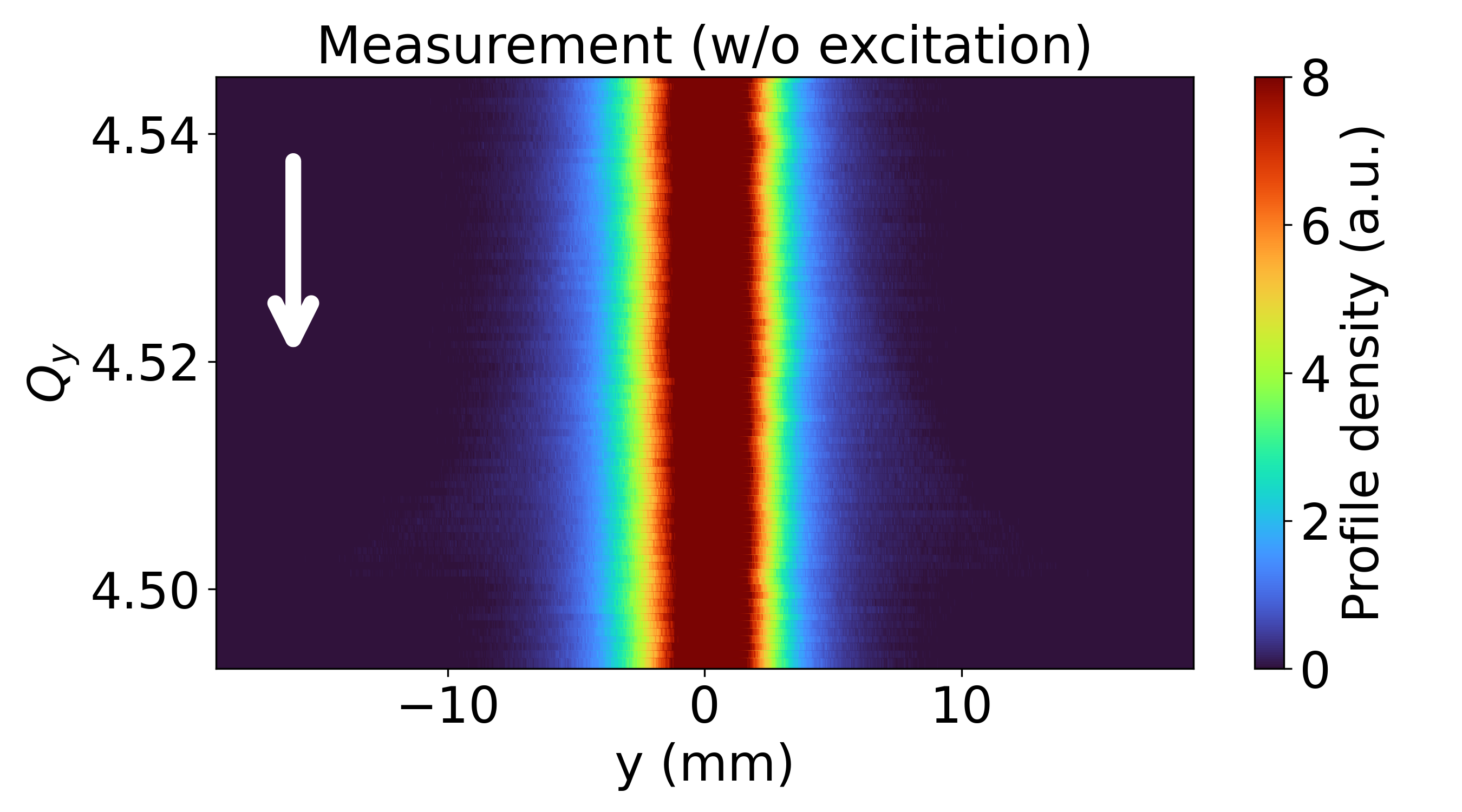}
   \includegraphics*[width=.99\columnwidth]{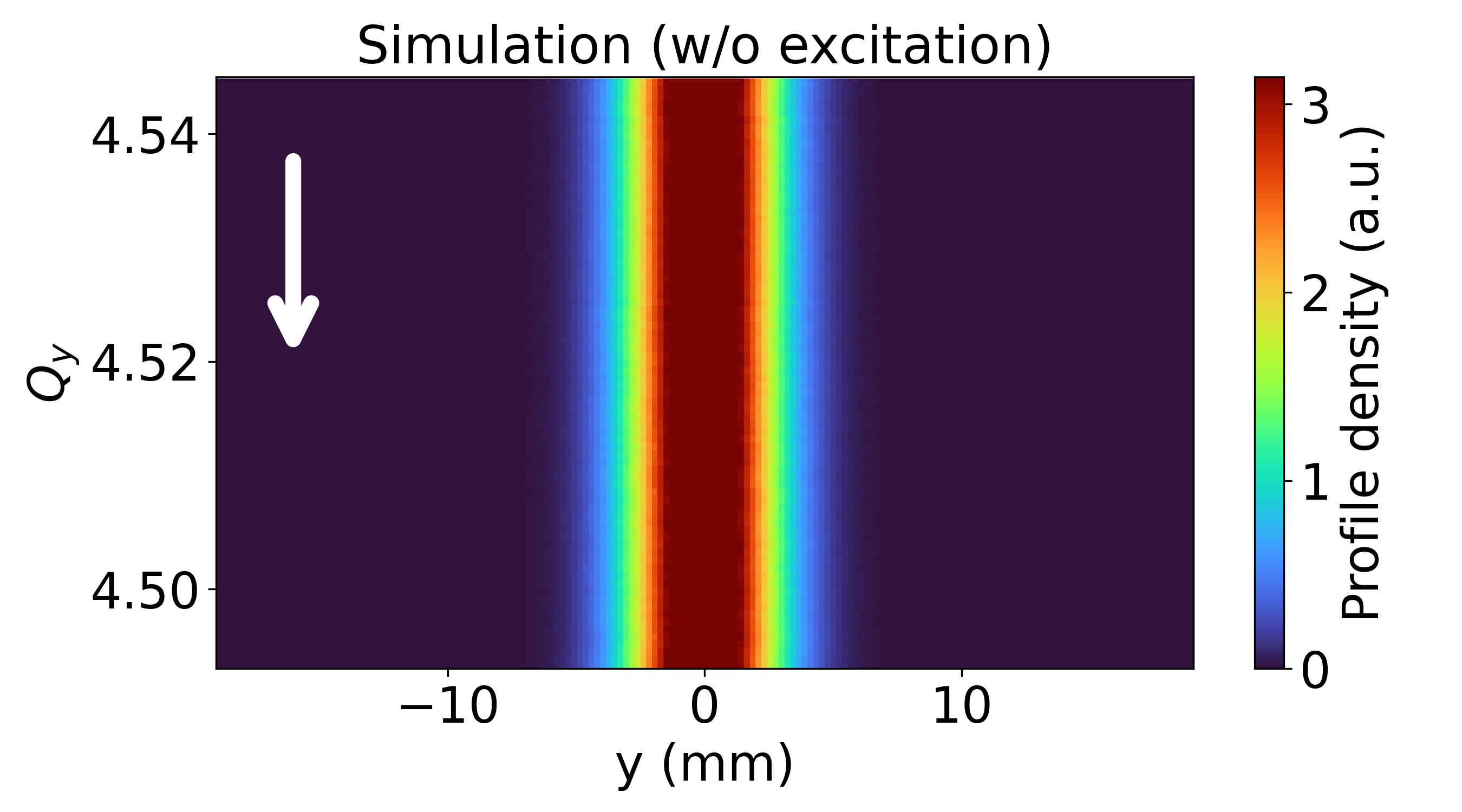}
   \caption{Top: evolution of the measured vertical beam profile as a function of the vertical tune during the crossing of the \textbf{compensated} $2Q_y=9$ resonance (white arrow points the direction of crossing). The vertical axis corresponds to the measured tune, the horizontal axis to the $y$-position along the beam profile, and the color to the density of the profile in arbitrary units. Bottom: simulation of the same process using a 6D tracking simulation.}
   \label{fig:no-trapping}
\end{figure}

The effect of resonance crossing on the beam was studied via vertical beam profile measurements acquired with a wire scanner~\cite{Veness:2289486} every $\sim$~1000 turns (corresponding to 1~ms beam storage time). Figure~\ref{fig:no-trapping} (top) shows the beam profiles of 150 PSB cycles, with the same beam and machine conditions, but taken at different moments during the cycle. Although there are some unavoidable fluctuations in the injection efficiency and beam intensity from cycle to cycle, these effects are within the noise of the profile measurements. The same technique is used for measuring the profile evolution during the different crossing scenarios.

As shown in Fig.~\ref{fig:no-trapping} (top), the vertical beam profiles remain unaffected from the resonance during the crossing. No beam loss or vertical emittance increase was measured.

This experimental setup was also studied with 6D tracking simulations using the same beam and machine parameters (Table~\ref{tab:simulation_params_crossing_above} of Appendix~\ref{sec:experimental_setup}). The simulated vertical beam profiles, shown in Fig.~\ref{fig:no-trapping} (bottom), showed no losses and no change in the vertical beam size or vertical emittance. 

The absence of any measurable effect on the beam in the experiment (and simulations) confirms that the half-integer resonance $2Q_y=9$ is not driven by space charge in the PSB (due to the machine periodicity of 16) and that the experimental resonance compensation sufficiently suppresses the resonance driving terms caused by lattice imperfections. 

\subsection{Particle trapping in resonance islands}\label{subsec:particle_trapping}

The same resonance crossing shown in Fig.~\ref{fig:tune_ramp} was repeated, this time with the resonance externally excited (as described in Sec.~\ref{sec:half-integer_resonance_control}) to produce a stopband width of $\delta Q_{res} \approx 0.004$. All other beam and machine parameters were kept identical to those of Table~\ref{tab:beam_params_crossing_above} of Appendix~\ref{sec:experimental_setup}. 

\begin{figure}[!htb]
   \centering
   \includegraphics*[width=.99\columnwidth]{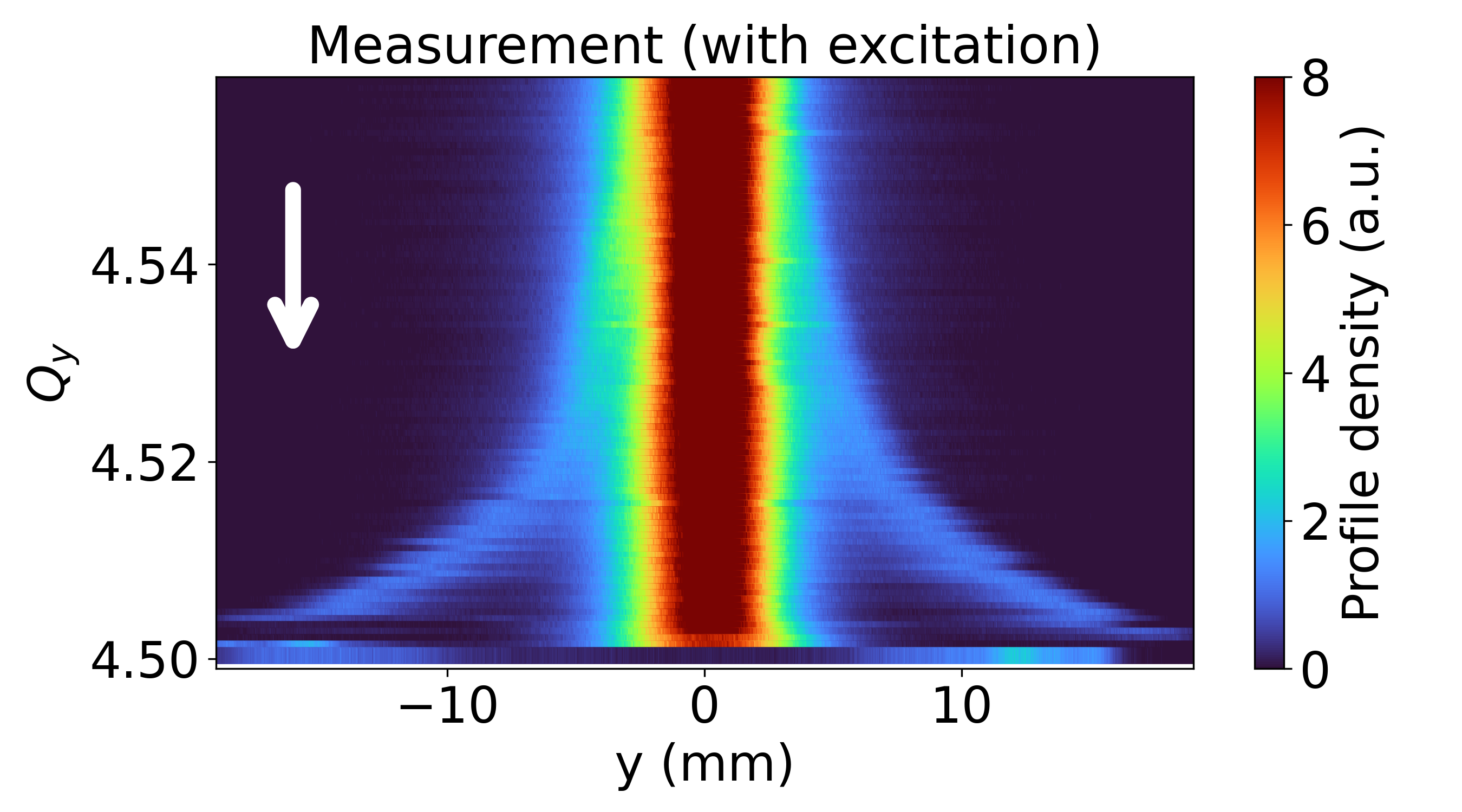}
   \includegraphics*[width=.99\columnwidth]{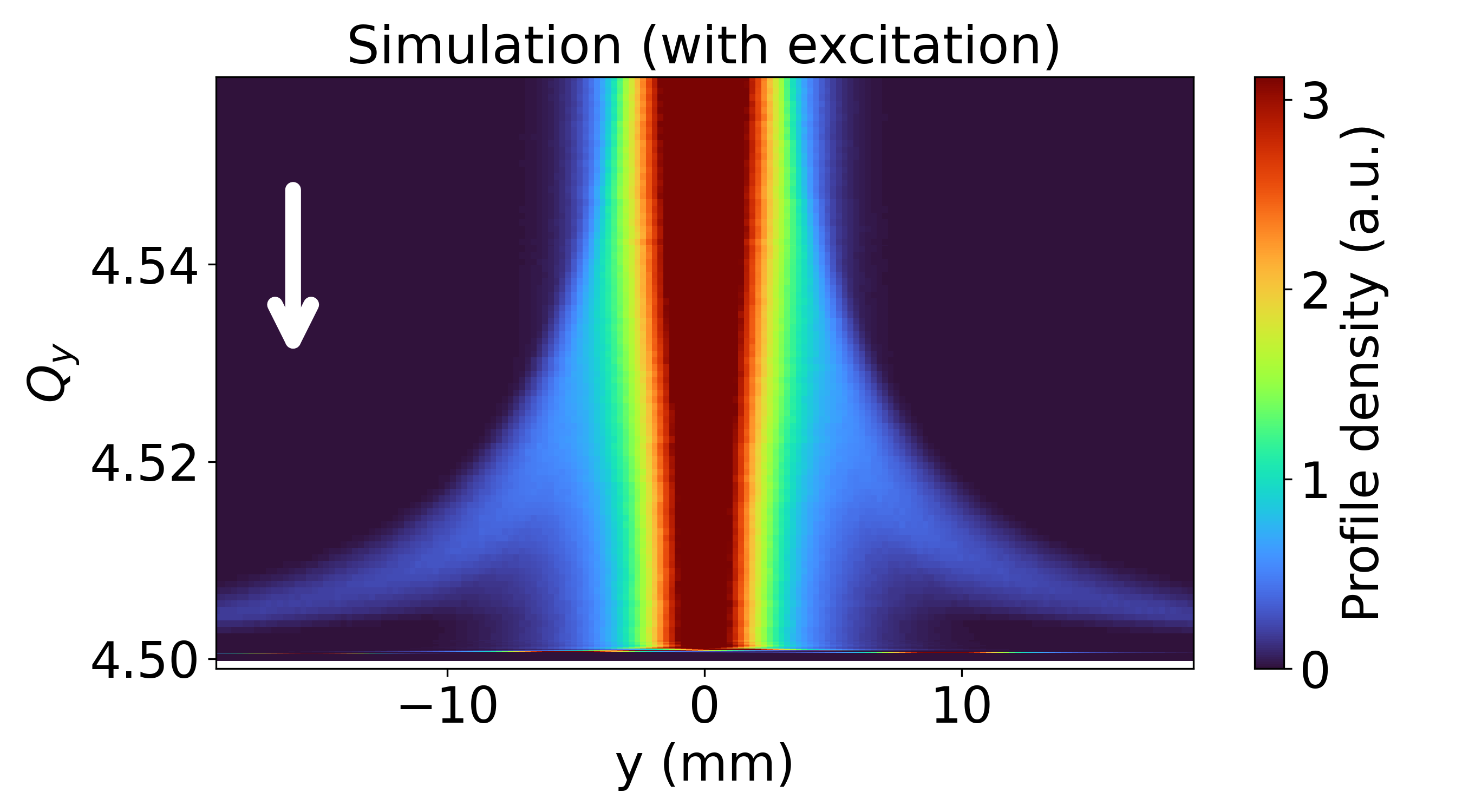}
   \caption{Top: evolution of the measured vertical beam profile as a function of the vertical tune during the crossing of the \textbf{excited} $2Q_y=9$ resonance (white arrow points the direction of crossing). Bottom: simulation of the same process using a 6D tracking simulation.}
   \label{fig:vertical_profile_evolution}
\end{figure} 

The measured evolution of the vertical beam profile is shown in Fig.~\ref{fig:vertical_profile_evolution} (top). Far from the resonance, the beam profiles are approximately Gaussian. As the vertical tune approaches the half-integer resonance, two regions around the beam core become populated with particles, forming two beamlets. These beamlets move outwards as the tune moves closer to $Q_y = 4.5$, until they reach the machine aperture and are lost (at $Q_y \approx 4.503$). Notably, the beam core is also lost shortly after. The reasons for that are discussed in Sec.~\ref{subsec:centroid_oscillations}.

As before, the experimental setup was studied also with 6D tracking simulations. The simulated vertical beam profiles, shown in Fig.~\ref{fig:vertical_profile_evolution} (bottom), match very well the experimental data, both in the formation and the movement of the beamlets. 

\begin{figure}[!htb]
  \centering
  \makebox[\linewidth][l]{\hspace*{3.5mm}\includegraphics[width=.77\linewidth]{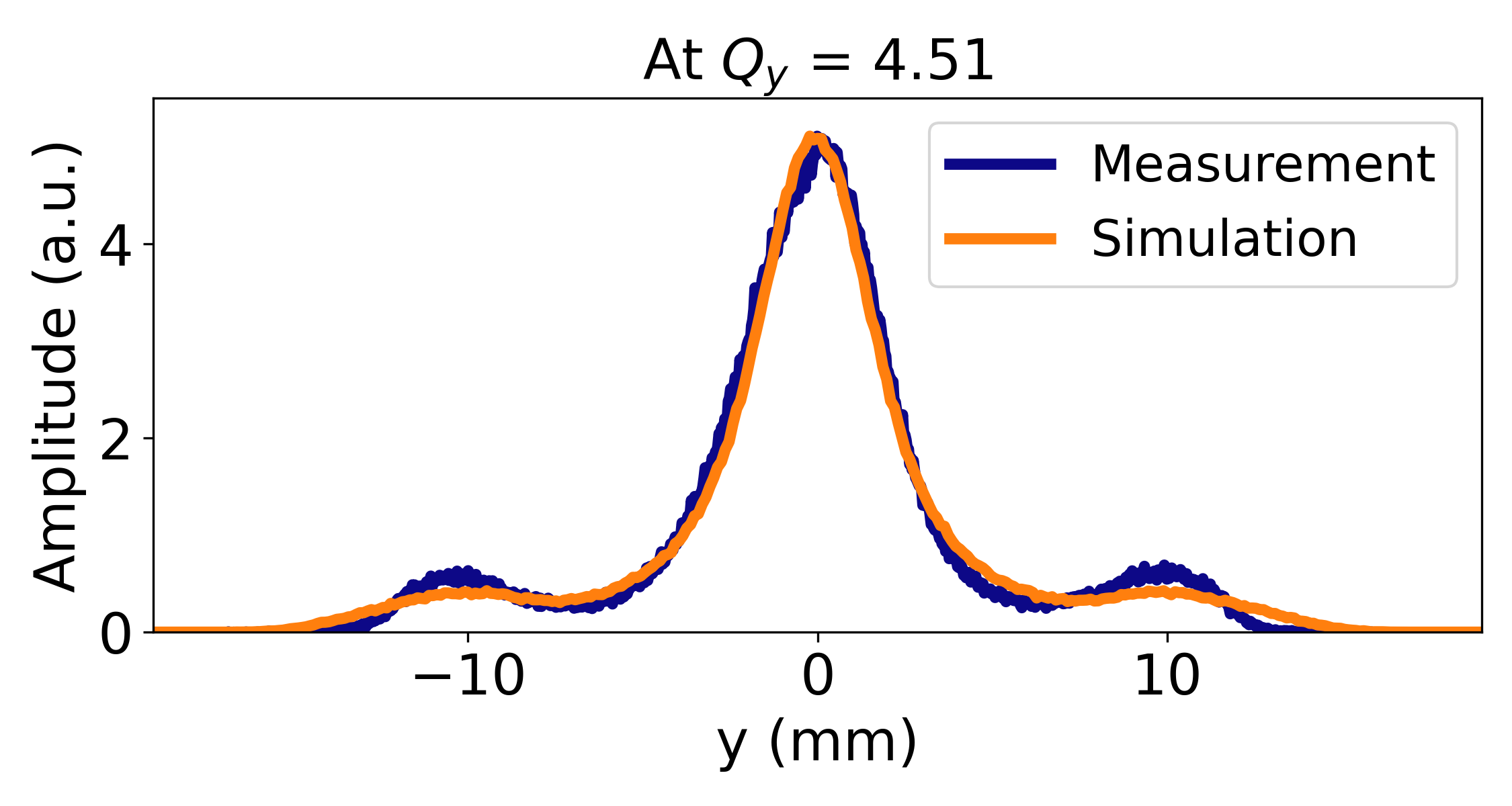}}
  \includegraphics[width=.99\linewidth]{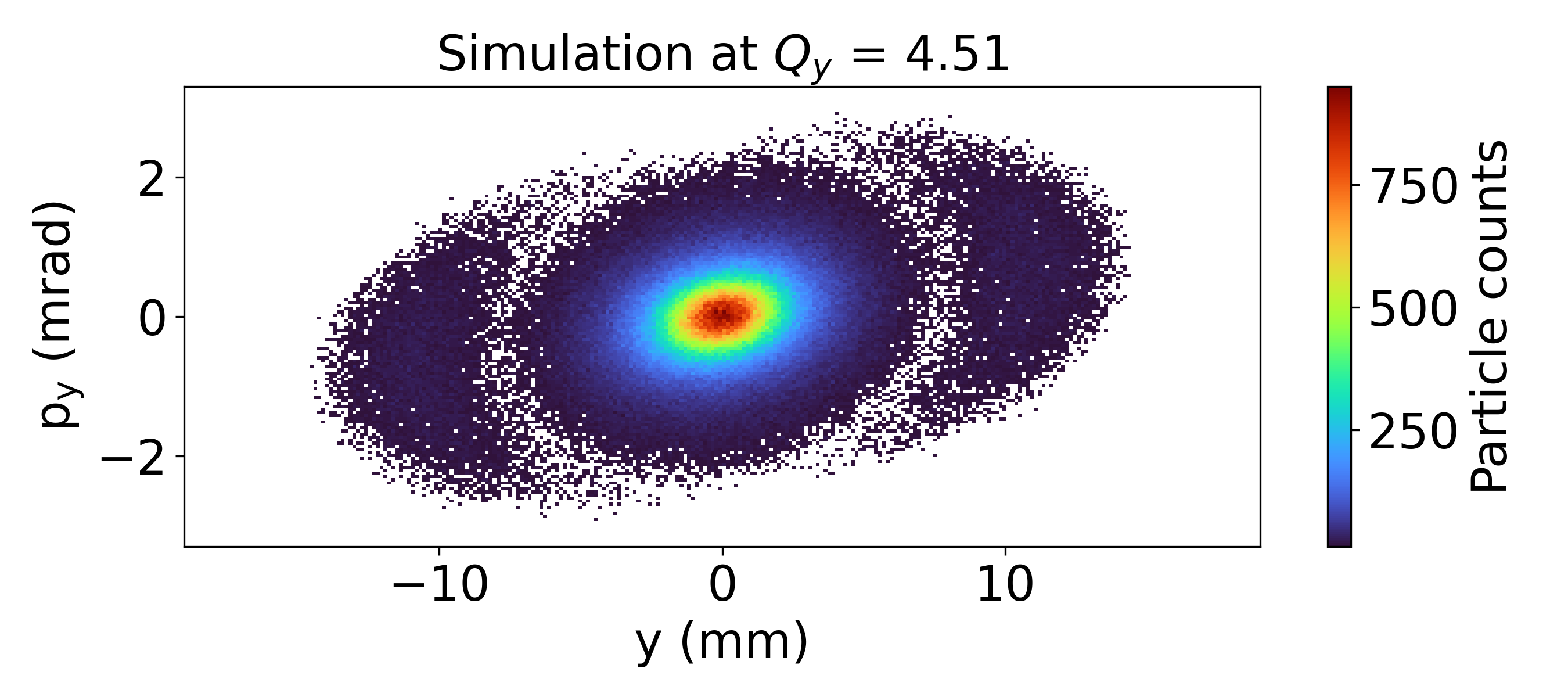}
  \caption{Top: Measured (blue) and simulated (orange) vertical beam profile during the crossing of the half-integer resonance at $Q_y=4.51$. Bottom: simulated vertical phase space at the same tune.}
  \label{fig:vertical_profile_single}
\end{figure}

By examining the vertical phase space evolution in the simulations it is observed that particles near the beam core get trapped in stable resonance islands due to the incoherent space charge as the vertical tune approaches the half-integer resonance, which is consistent with previous studies~\cite{FRANCHETTI2006195}. As the tune moves closer to $Q_y=4.5$, the islands, along with the trapped particles, move outward, forming the observed beamlets (as shown in Fig.~\ref{fig:vertical_profile_single}). When the resonance islands reach the machine aperture, the trapped particles are lost. The beam core is also lost in the simulation.

The agreement between measurement and simulation supports the interpretation that the beam evolution during the resonance crossing is driven by particle trapping in the resonance islands due to the incoherent space charge. No oscillations of the vertical beam envelope were observed within the studied parameter range, neither in measurements nor simulations. Adjusting the resonance excitation allows controlling the island orientation, and thus the position and visibility of the beamlets in the projection of the phase space onto the beam profile, as shown experimentally in Appendix~\ref{appendix:orientation_of_beamlets}. An experiment demonstrating the return of trapped particles towards the beam core using a reversed tune ramp is presented in Appendix~\ref{appendix:remerging}.

\FloatBarrier 

\subsection{Centroid oscillations during resonance crossing}\label{subsec:centroid_oscillations}

\begin{figure}[!htb]
   \centering
   \includegraphics*[width=.99\columnwidth]{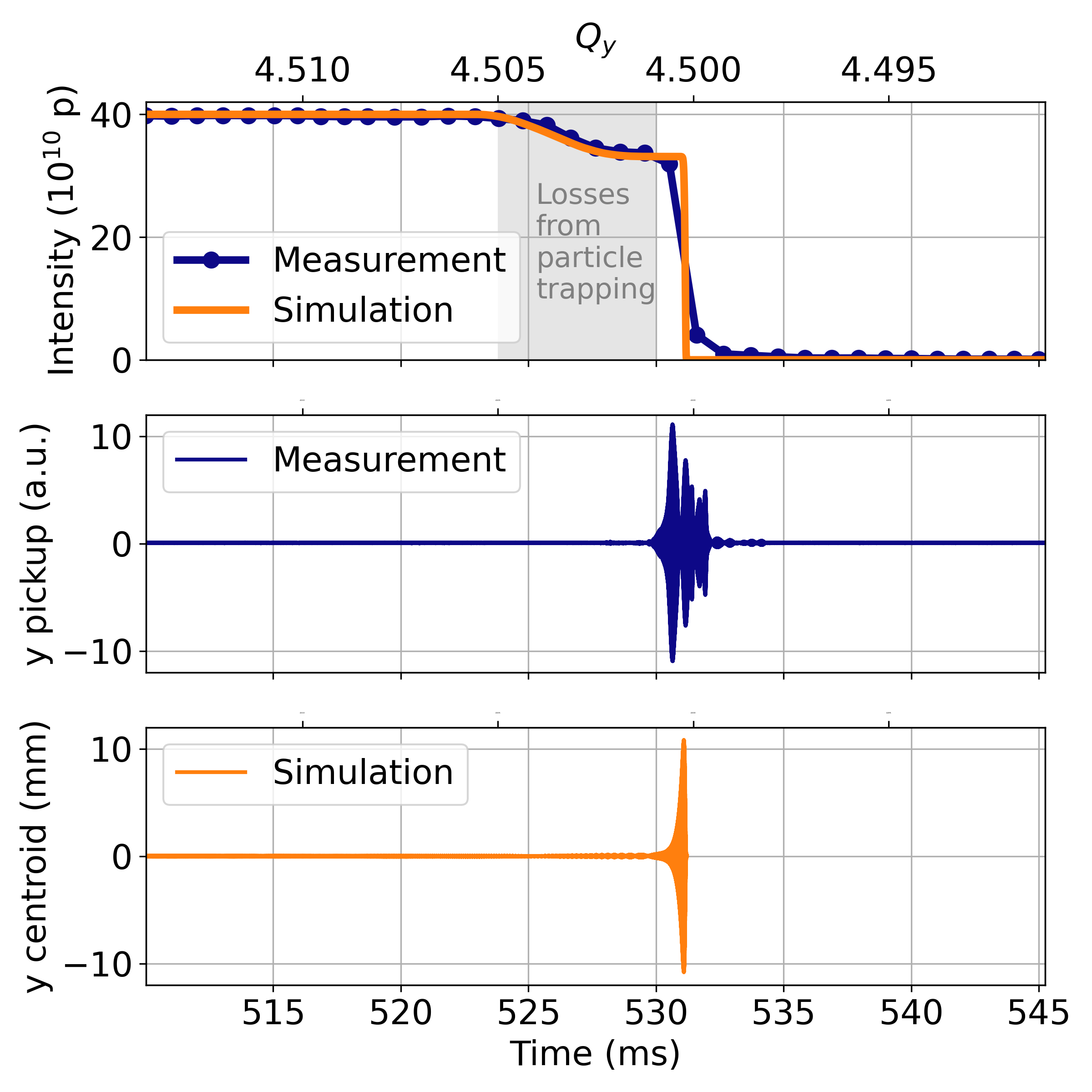}
   \caption{Top: measured (blue) and simulated (orange) beam intensity as a function of time during the crossing described in Fig~\ref{fig:tune_ramp} ($\delta Q_y / \text{turn} = -0.66 \times 10^{-6}$). The beam loss for $Q_y$ in the interval $[4.505, 4.501]$ (shaded area) is associated with the beamlets (i.e.~particles trapped in the resonance islands) intercepting the machine aperture. The loss of the rest of the particles takes place for $Q_y$ in the interval $[4.501, 4.499]$. Middle: measurement of the vertical beam centroid as a function of time using a turn-by-turn signal (growth rate of $\lambda = 0.005$ turn$^{-1}$). Bottom: vertical beam centroid as a function of time from tracking simulations (similar growth rate).}
   \label{fig:intensity_tbt_measurement_simulation}
\end{figure}

The beam intensity evolution during the half-integer resonance crossing is consistent between measurement and simulation. As shown in Fig.~\ref{fig:intensity_tbt_measurement_simulation} (top), the loss of the trapped particles occurs first (shaded area), followed by a fast loss of the remaining beam particles. Just before the fast beam loss starts, vertical beam centroid oscillations were observed in the measurements using a turn-by-turn pickup~\cite{Gasior:883298,Gasior:2019hxg} as shown in Fig.\ref{fig:intensity_tbt_measurement_simulation} (middle). These oscillations were reproduced in the tracking simulations as shown in Fig.~\ref{fig:intensity_tbt_measurement_simulation} (bottom), and further confirmed independently with wire scanner measurements, as discussed in Appendix~\ref{appendix:wire-scattering}.

To investigate the origin of the beam centroid oscillations, additional tracking simulations were performed. First, a single particle was tracked in the PSB lattice with a constant vertical tune inside the half-integer resonance stopband. Turn-by-turn, an exponential growth of the particle’s vertical action was observed:
\begin{equation}
    J_y \propto e^{\lambda n},
\end{equation}
where $n$ is the turn number and $\lambda$ is the growth rate. 

\begin{figure}[htb]
   \centering
   \includegraphics*[width=.99\columnwidth]{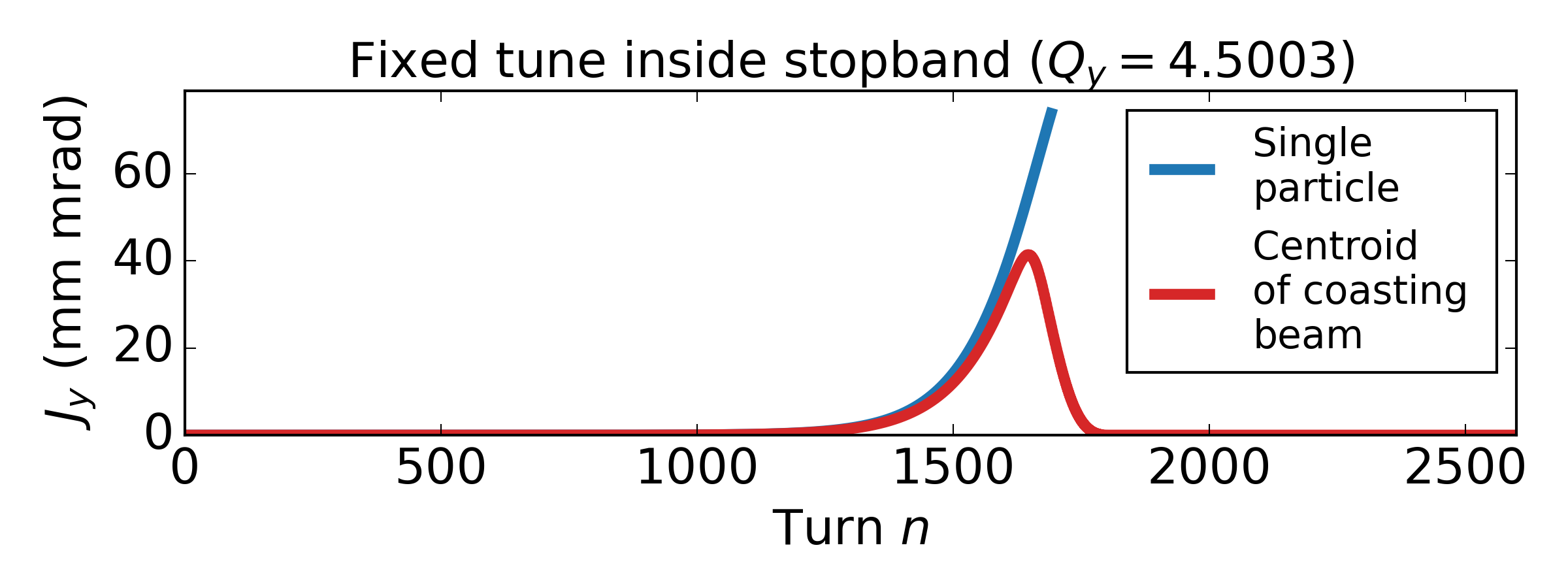}
   \includegraphics*[width=.99\columnwidth]{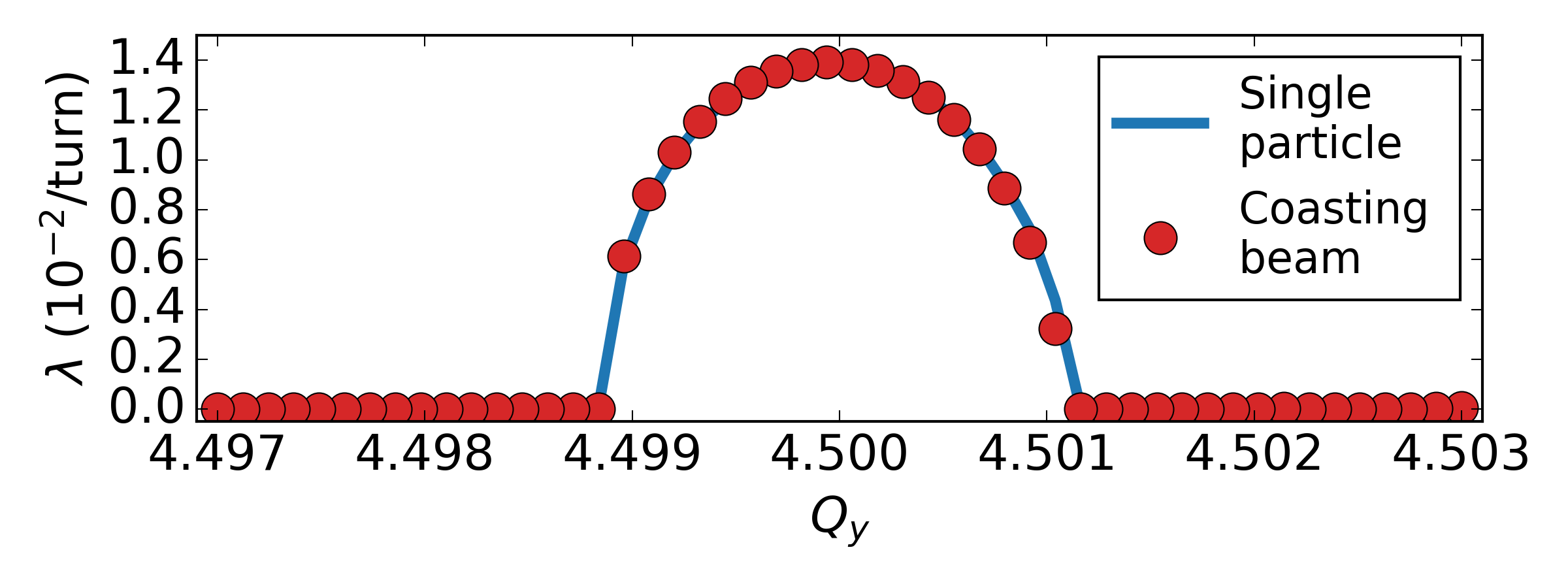}
   \caption{Top: simulation of the turn-by-turn vertical action of a single particle (blue) and the beam centroid of a coasting beam with space charge (red) for a fixed tune inside the half-integer resonance stopband ($Q_y=4.5003$). Bottom: simulated growth rate $\lambda$ of the single particle vertical action (blue) and coasting beam vertical centroid action (red) as a function of the vertical tune.}
   \label{fig:coherent_dipolar_oscillations_simulations}
\end{figure}

Then, a simulation was performed using a coasting beam, with a constant vertical tune set inside the stopband (as before). Space charge was included and was sufficiently large to shift the individual particle tunes below the stopband. Since the center-of-mass does not feel the space charge~\cite{10.1063/1.56781} (as the forces net out), it effectively behaves like a single particle and thus vertical centroid oscillations develop, and the centroid action also exponentially increases, as shown in Fig.~\ref{fig:coherent_dipolar_oscillations_simulations} (top).

The growth rates of the vertical centroid action (for the coasting beam with space charge) and of the single-particle action were compared for a range of vertical tunes, both inside and outside the stopband. As shown in Fig.~\ref{fig:coherent_dipolar_oscillations_simulations} (bottom), the two growth rates are identical, including when the machine tune (i.e. center-of-mass tune) is within the resonance stopband. This suggests that, when space charge is strong enough to shift the individual particle tunes far away from the stopband, the beam centroid behaves effectively like a single particle and becomes unstable when the machine tune lies within the stopband.

In the crossing of the half-integer resonance of Fig.~\ref{fig:intensity_tbt_measurement_simulation}, space charge calculations confirm that, even after the initial loss of trapped particles, the remaining beam has still strong enough space charge to shift the individual particle tunes well below the resonance. However, while the machine tune continues to ramp, the center-of-mass tune still enters the resonance stopband, similar to the situation of the simulated static tune scan shown in Fig.~\ref{fig:coherent_dipolar_oscillations_simulations}. If the resonance crossing is slow enough, the center-of-mass tune may spend a sufficient amount of time within the stopband for centroid oscillations to develop and grow, eventually leading to the observed fast loss. 

\subsection{Scaling of beam losses}\label{subsec:scaling_beam_losses}

Beam losses during the half-integer resonance crossing, either associated with the particle trapping or the centroid oscillations, depend on the resonance crossing rate, the maximum space charge tune shift, and the resonance stopband width. To study how losses scale with these parameters, a series of systematic measurements and simulations were carried out. The crossing rate was adjusted by modifying the quadrupole ramp rate, while space charge was controlled by changing the injected beam intensity while keeping the transverse emittances relatively constant. In all measurements, the resonance stopband width was kept fixed, and the total beam loss after the resonance crossing was recorded.

\begin{figure}[!htb]
   \centering
   \includegraphics*[width=.99\columnwidth]{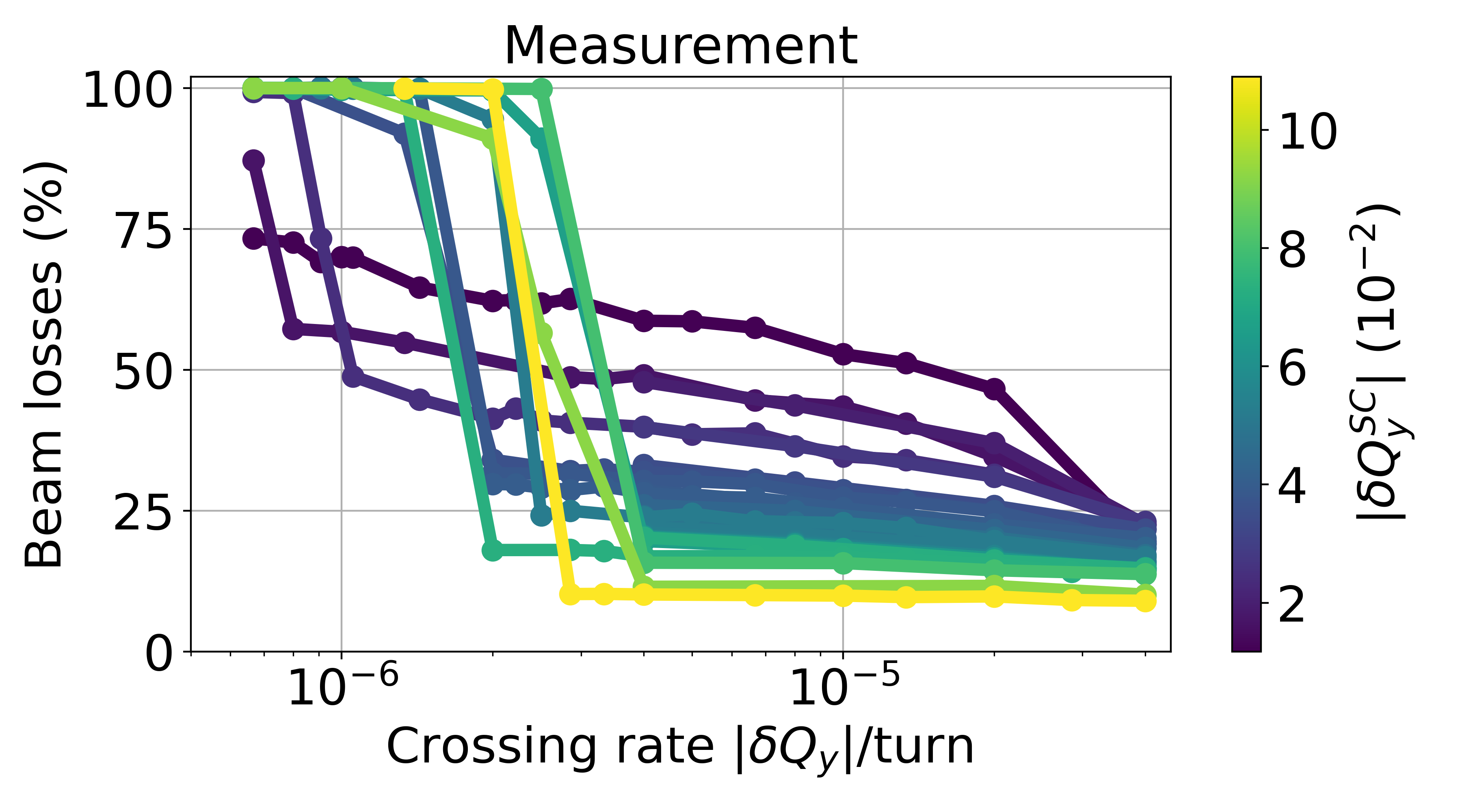}
   \includegraphics*[width=.99\columnwidth]{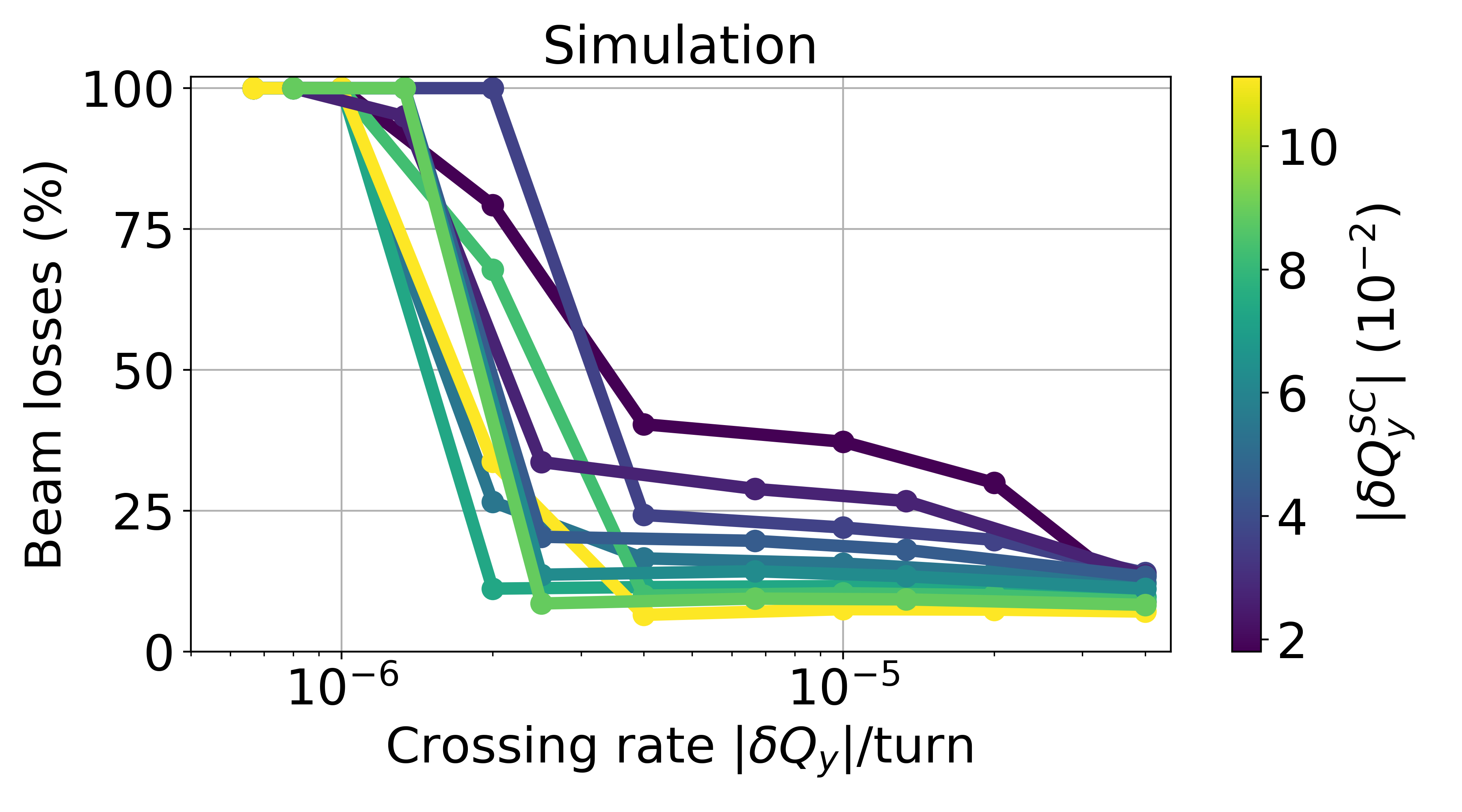}
   \caption{Top: measured beam losses during the half-integer resonance crossing from above for different crossing rates and maximum space charge tune shift. The half-integer resonance stopband width is fixed ($\delta Q_{y,\textrm{res}} \approx 0.004$).
   Bottom: beam losses from tracking simulations using the same beam and machine parameters as above.}
   \label{fig:crossing_speed_multiple_spacecharge}
\end{figure}

As shown in the measurements of Fig.~\ref{fig:crossing_speed_multiple_spacecharge} (top), faster crossing rates and stronger space charge resulted in lower losses. In contrast, lower crossing rates often led to full beam loss, as illustrated also in Fig.~\ref{fig:intensity_tbt_measurement_simulation}.

To interpret these observations, 6D tracking simulations were performed for almost all combinations of crossing rate and space charge tune shift. The results, shown in Fig.~\ref{fig:crossing_speed_multiple_spacecharge} (bottom), closely reproduce the measurements: faster crossing rates and stronger space charge reduces the particle trapping in the resonance islands and lead to fewer losses. For slower crossing rates, the simulations show the growth of the centroid oscillations which leads to the complete beam loss. The agreement between simulation and measurements across all cases supports the interpretation of the loss mechanisms described in Secs.~\ref{subsec:particle_trapping}~and~\ref{subsec:centroid_oscillations}.

\begin{figure}[!htb]
   \centering
   \includegraphics*[width=.99\columnwidth]{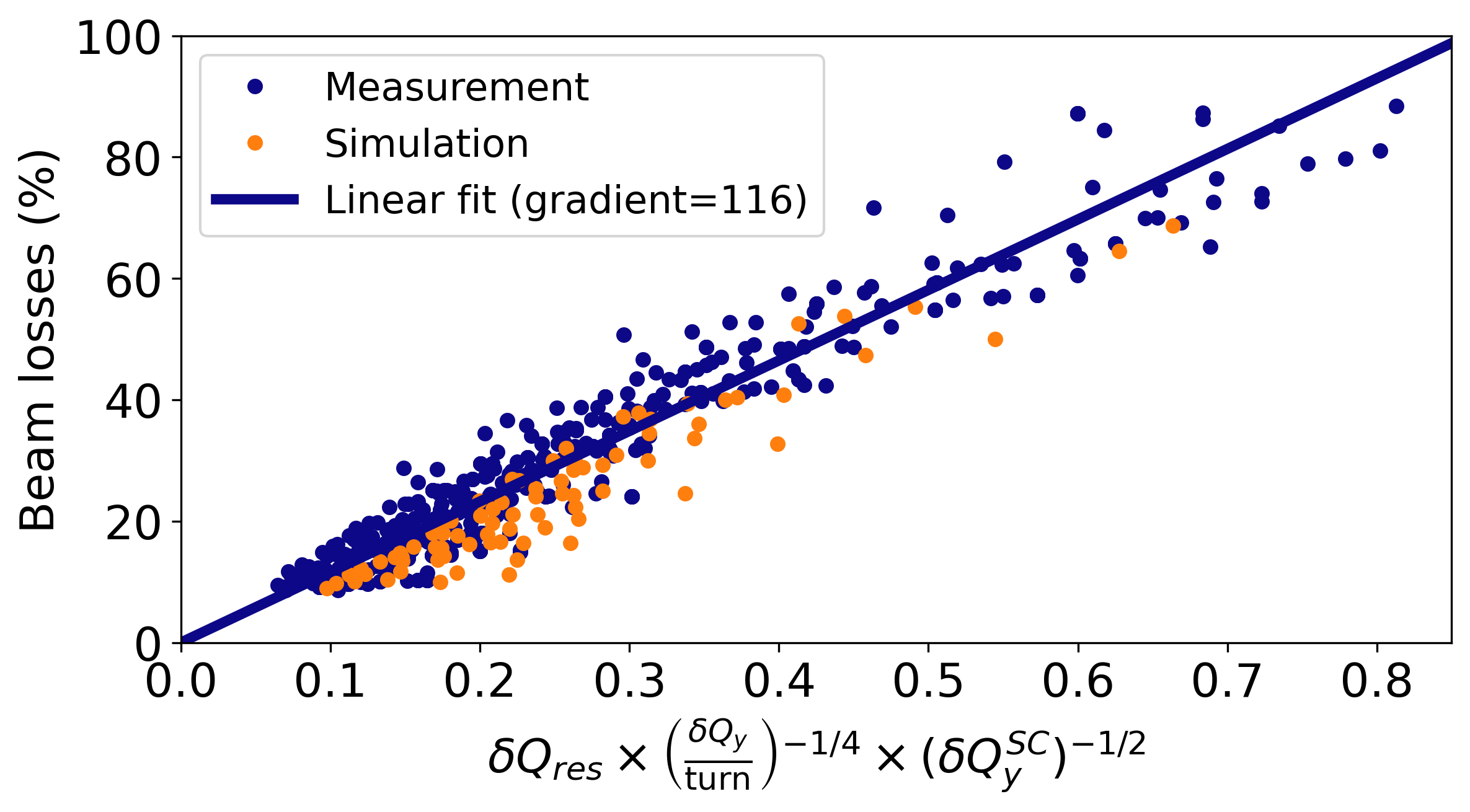}
   \caption{Experimental data (blue points) and linear fit (blue line) on the beam losses after crossing the half-integer resonance from above with a coasting beam using different values of the crossing rate $\delta Q_y / \mathrm{turn}$, maximum space charge tune shift $\delta Q_y^{SC}$ (in absolute value), and resonance stopband width $\delta Q_{y,\textrm{res}}$. Data from tracking simulations with similar parameters are included (orange points).}
   \label{fig:scaling_final}
\end{figure}

The dependence on the resonance stopband width was explored by varying the resonance excitation in both measurements and simulations. Across all parameter scans, the measured and simulated losses are found to follow the empirical scaling:
\begin{equation} \label{eq:scaling_final}
    L \propto \delta Q_{y,\mathrm{res}} \times \left(\frac{\delta Q_y}{\mathrm{turn}}\right)^{-\frac{1}{4}} \times \left| \delta Q_y^{SC} \right| ^{-\frac{1}{2}},
\end{equation}
where $L$ is the total beam loss, $\delta Q_{y,\textrm{res}}$ is the resonance stopband width, $\delta Q_y / \mathrm{turn}$ is the (absolute) crossing rate, and  $\delta Q_y^{SC}$ is the (absolute) maximum space charge tune shift. This scaling is illustrated in Fig.~\ref{fig:scaling_final}. Data points corresponding to full beam losses were excluded, so the scaling describes only the losses from trapped particles.  

\section{Half-Integer Crossing from Below the Resonance}\label{sec:half-integer_resonance_crossing_below}

The half-integer resonance was also crossed from below the resonance, i.e.~from $Q_y<4.5$ to $Q_y>4.5$, with a crossing rate of $\delta Q_y / \text{turn} = +0.66 \times 10^{-6}$. As before, the resonance stopband was $\delta Q_{y,\textrm{res}} \approx 0.004$, a coasting beam with a maximum space charge tune shift of $\delta Q_y^{SC} \approx -0.09$ was used, and the vertical chromaticity was compensated (similar setup as in Table~\ref{tab:beam_params_crossing_above}). 

\begin{figure}[!htb]
   \centering
   \includegraphics*[width=.99\columnwidth]{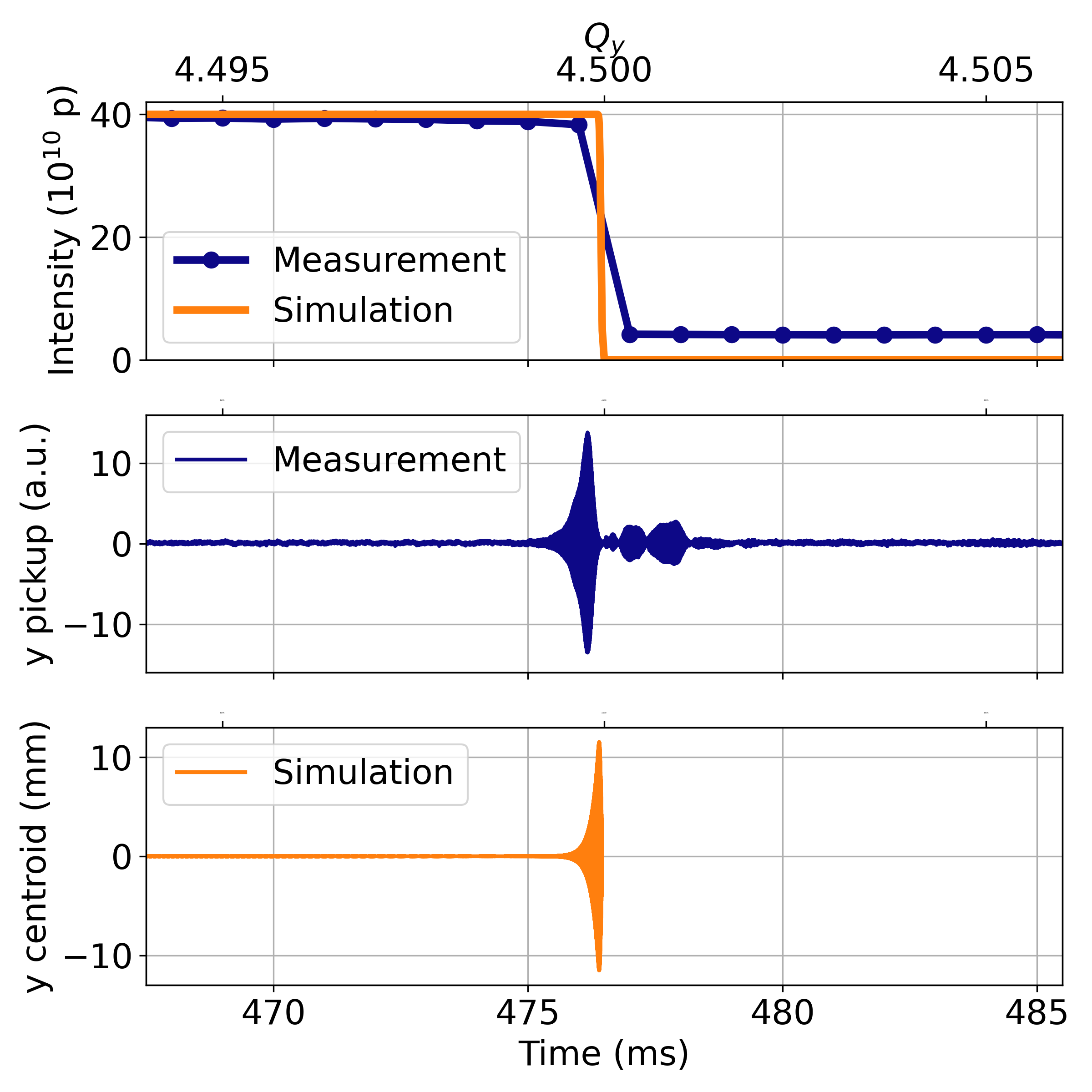}
   \caption{Top: measured (blue) and simulated (orange) beam intensity as a function of the tune during the half-integer resonance crossing from below ($\delta Q_y / \text{turn} = +0.66 \times 10^{-6}$). Middle: measurement of the vertical beam centroid as a function of the tune using a turn-by-turn pickup. Bottom: vertical beam centroid as a function of the tune from tracking simulations. }
   \label{fig:crossing_below_instability}
\end{figure}

As shown in Fig.~\ref{fig:crossing_below_instability}, during this crossing almost the entire beam is lost, while vertical centroid oscillations are observed with a turn-by-turn pickup. Both the losses and the oscillations are well reproduced in simulations. In simulations, the vertical centroid motion grows when the center-of-mass tune enters the stopband, while the space charge tune shift remains approximately unchanged. The mechanism described in Sec.~\ref{subsec:centroid_oscillations} can provide also here an explanation for the experimental observations. 

A total beam loss can be avoided by increasing the crossing rate. At three times the original rate (i.e.~$\delta Q_y / \text{turn} = +2 \times 10^{-6}$), almost no beam loss and no significant centroid oscillations were observed. This allowed for the study of the beam dynamics during the crossing from below using vertical profile measurements.

\begin{figure}[!htb]
   \centering
   \includegraphics*[width=.99\columnwidth]{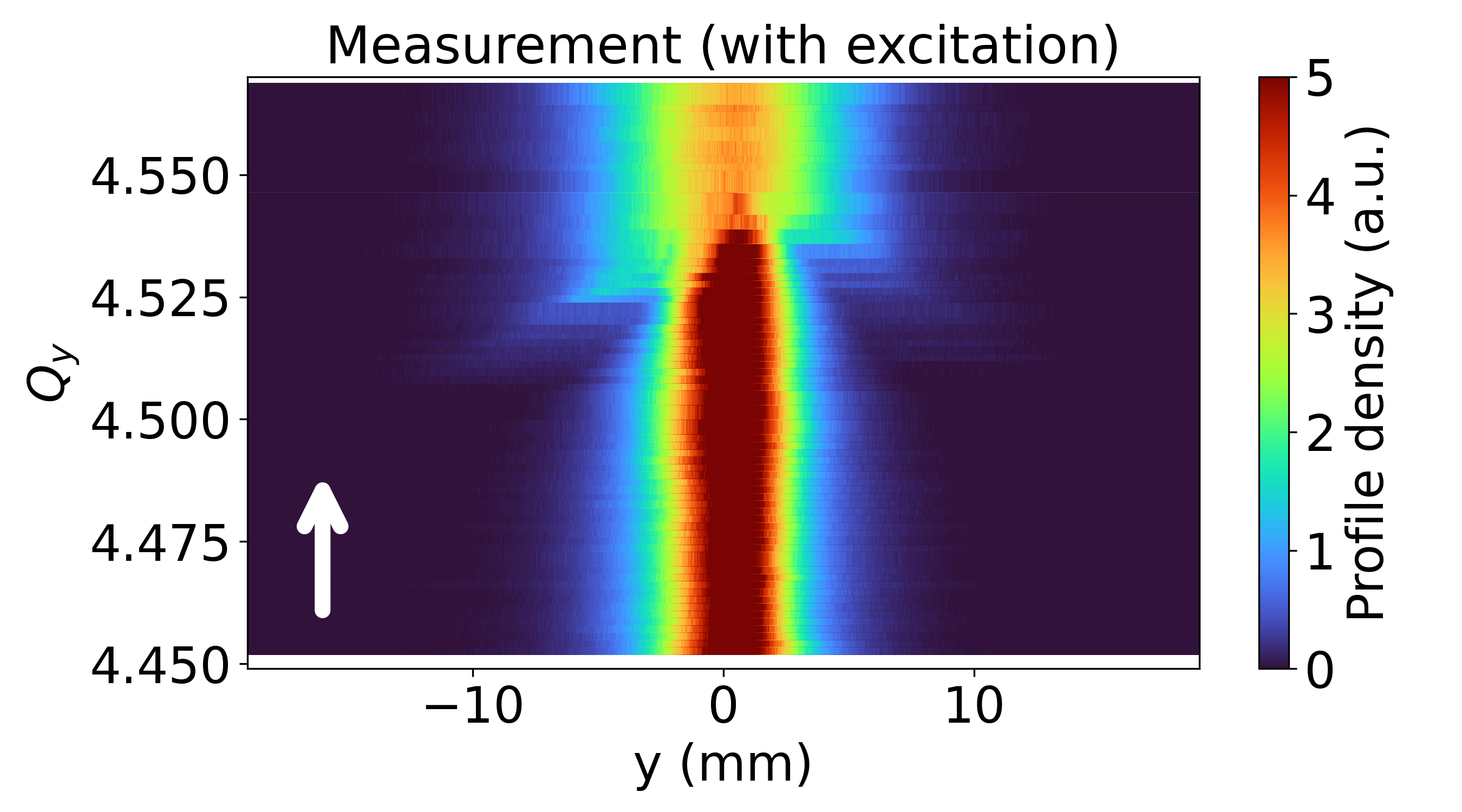}
   \includegraphics*[width=.99\columnwidth]{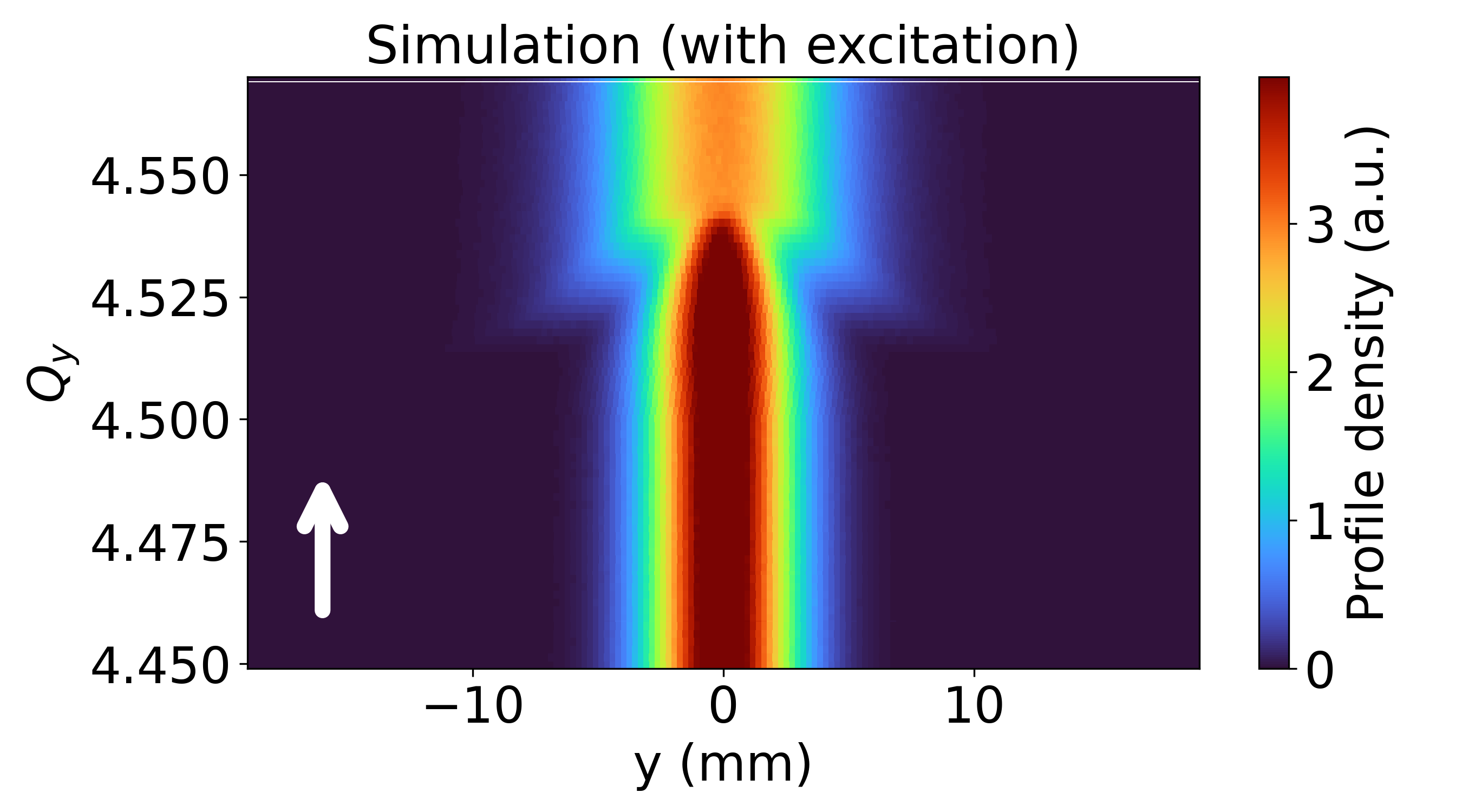}
   \caption{Top: evolution of the measured vertical beam profile as a function of the vertical tune during the crossing of the $2Q_y=9$ resonance from below with $\delta Q_y / \text{turn} = +2 \times 10^{-6}$ (white arrow points the direction of crossing). The vertical axis corresponds to the measured tune, the horizontal axis to the $y$-position along the beam profile, and the color to the density of the profile in arbitrary units. Bottom: simulation of the same process using a 6D tracking simulation.}
   \label{fig:crossing_below_full}
\end{figure}

Figure~\ref{fig:crossing_below_full} (top) shows around 100 measured vertical profiles as a function of the vertical tune (similarly to how it is done in Fig.~\ref{fig:vertical_profile_evolution}). Below the resonance (e.g.~$Q_y\approx 4.47$) the profiles are Gaussian. As the machine tune approaches the half-integer, particles begin populating the edge of the distribution. Eventually, the beam size grows and remains larger, as the tune continues to ramp, with minimal losses. 

\begin{figure*}[!hbt]
  \centering
  \includegraphics[width=.99\textwidth]{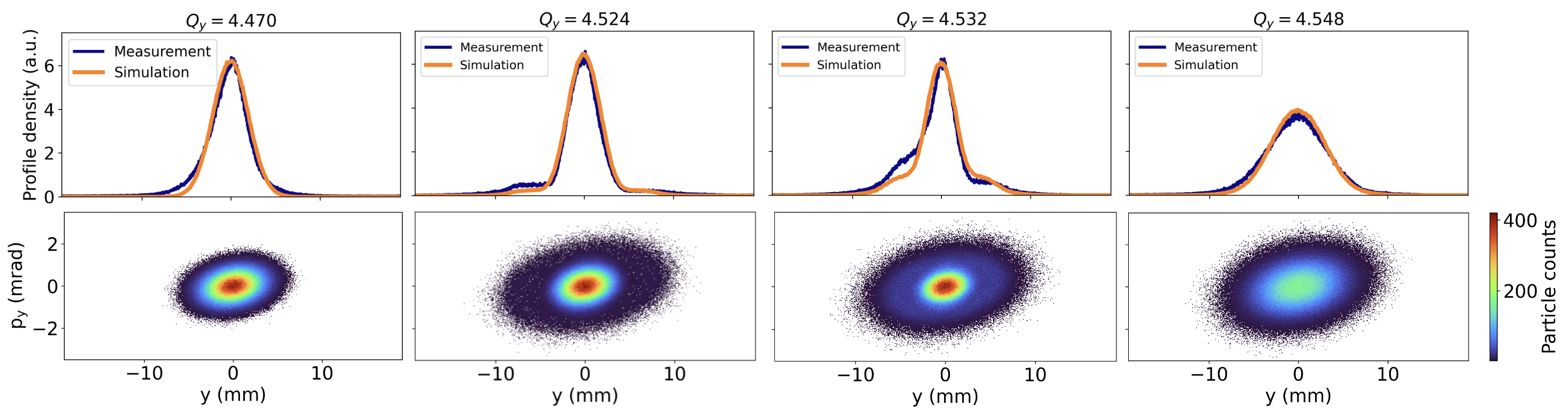}
  \caption{Top: measured (blue) and simulated (orange) vertical beam profiles at different tunes during the crossing of the half-integer resonance from below with $\delta Q_y / \text{turn} = +2 \times 10^{-6}$. Bottom: simulated vertical phase space at the same tunes as the measurements. The asymmetry observed in the measured beam profiles is an artifact introduced by the measurement device, as discussed in Appendix~\ref{appendix:wire-scattering}.}
  \label{fig:crossing_below_profiles}
\end{figure*}

This setup was also reproduced using 6D tracking simulations with space charge. The full evolution of the simulated vertical profiles are shown in Fig.~\ref{fig:crossing_below_full} (bottom) while individual profiles and phase spaces are shown in Fig.~\ref{fig:crossing_below_profiles}. When crossing from below, the resonance islands move inward toward the beam center. As the resonance islands reach the edge of the distribution, particles can cross the separatrix and move to larger amplitudes (e.g.~at $Q_y\approx 4.532$). As the tune ramp continues, the resonance islands shrink and collapse towards the center of the phase space, resulting in a larger phase space distribution.

The similarity between the measured profiles and the simulated phase space distributions supports an interpretation in which the beam growth in the vertical plane is driven by the inwards motion of the resonance islands created due to the incoherent tune spread from space charge. The asymmetries observed in the measured profiles are introduced by the measuring instrument, and are discussed in Appendix~\ref{appendix:wire-scattering}. 

For several resonance crossing rates, space charge tune shifts, and stopband widths, the beam losses during the resonance crossing from below were either close to 100\% (as in Fig.~\ref{fig:crossing_below_instability}) or very small (as in Fig.~\ref{fig:crossing_below_full}), so no scaling applies.

%\FloatBarrier

\section{Conclusions}\label{sec:conclusions_outlook}

In this paper, the crossing of a controlled non-structure half-integer resonance $2Q_y=9$ in the PSB was carried out with a coasting beam. The experimental results have been compared to 6D tracking simulations of the same scenarios. This study demonstrates experimentally the role of the incoherent space charge driven effects in the half-integer resonance crossings and provides quantitative benchmarks against detailed simulations. 

For resonance crossings from above the resonance, the close agreement between measurement and simulation supports the interpretation that the beam evolution is primarily affected by particle trapping in resonance islands induced by the incoherent space charge. At slow crossing rates, oscillations of the beam centroid were observed, while no coherent envelope oscillations were measured (nor observed in the corresponding tracking simulations). Systematic scans revealed an empirical scaling law for beam losses as a function of resonance crossing rate, space charge tune spread, and resonance strength, again with very good agreement between measurements and simulations.

For resonance crossings from below the resonance, the agreement between measured profiles and simulated phase-space distributions indicates that the observed vertical beam growth is driven by the inwards motion of the resonance islands, which is defined by the incoherent space charge. As in the case of resonance crossing from above, centroid oscillations were observed at slow crossing rates, which ultimately lead to complete beam loss.

These results are relevant for the PSB and other high-intensity, high-brightness synchrotrons operating close to half-integer tunes. In particular for the PSB, an injection above the half-integer resonance and its fast crossing emerges as a potential mitigation strategy against space charge induced emittance blow-up at low energy for increasing the brightness performance reach. Studies in this direction will be part of future work.

\begin{acknowledgments}
    The authors would like to thank S.~Albright, G.~P.~Di Giovanni, P.~D.~Hermes, T.~Levens, M.~Giovannozzi, E.~Métral, Y.~Papaphilippou, and F.~Roncarolo for their valuable inputs and discussions, G.~Iadarola and his team for providing \texttt{Xsuite} and their support, and the PSB Operations team for their assistance during the MD measurements. 
\end{acknowledgments}

%\FloatBarrier

\appendix

\section{\label{sec:experimental_setup}Details of the experimental and simulation setup}

All experiments were performed in a non-accelerating cycle at a kinetic energy of $160$~MeV (PSB injection energy). The coasting beam used in these experiments was formed by injecting a $600$~ns pulse from Linac4 and keeping all RF cavities switched off, letting it fully debunch after about $500$ turns ($\sim 0.5$~ms) after injection. 

The beam intensity was primarily controlled by varying the number of injected Linac4 pulses, while finer adjustments were made by slightly changing the pulse length. The transverse emittances were changed by varying the number of turns the beam went through the injection foil of the charge-exchange injection and also by missteering the injection trajectory~\cite{Renner_PhD}. By combining different intensities and emittances, the desired maximum space charge tune shifts were achieved.

The natural normalized chromaticities for the working point of $(Q_x,Q_y)=(4.45,4.15)$ were measured at $(\xi_x,\xi_y)=(Q'_x/Q_x,Q'_y/Q_y)=(-0.84,-1.60)$. Due to the limited number of sextupole correctors in the PSB, chromaticity correction is possible in only one plane at a time. For the experimental study presented here, only the vertical chromaticity was corrected close to zero, unavoidably increasing the horizontal chromaticity to $\xi_x \approx -1.59$. %The chromaticity correction setting was kept constant during the different resonance crossings as we checked that there is a minimal variation in the various tunes under study.

Table~\ref{tab:beam_params_crossing_above} lists the beam and machine parameters used in half-integer resonance crossings from above.

\begin{table}[!htp]
    \caption{Parameters used in the half-integer resonance crossings from above (Fig.~\ref{fig:no-trapping} and Fig.~\ref{fig:vertical_profile_evolution}).} 
    \centering
    \begin{tabular}{r l}
    \hline
    Beam intensity $N_b$ & $\approx 40 \times 10^{10}$~p \\
    Norm. hor. (Gaussian) emittance $\epsilon_x$ & $\approx 1.0$~mm mrad \\
    Norm. ver. (Gaussian) emittance $\epsilon_y$ & $\approx 0.63$~mm mrad\\
    Longitudinal shape: & coasting \\
    Relative momentum spread $(\delta p/p)_\textrm{RMS}$ & $\approx 1.4 \times 10^{-3}$ \\
    Max. hor. space charge tune shift $\delta Q_x^{SC}$ & $\approx -0.06$ \\
    Max. ver. space charge tune shift $\delta Q_y^{SC}$ & $\approx -0.09$ \\
    Initial working point $(Q_x, Q_y)$ & $= (4.15, 4.65)$ \\
    Final working point $(Q_x, Q_y)$ & $= (4.15, 4.45)$ \\
    Crossing rate $\delta Q_y / \text{turn}$ & $= -0.66 \times 10^{-6}$ \\
    Resonance stopband width $\delta Q_{y,\textrm{res}}$ & $\approx 0$ or $ 0.004$ \\
    Hor. normalized chromaticity $\xi_x$ & $\approx -1.59$ \\
    Ver. normalized chromaticity $\xi_y$ & $\approx 0$ \\
    \hline
    \end{tabular}  
    \label{tab:beam_params_crossing_above}
\end{table}

For the tracking simulations, the transverse particle distributions were Gaussian, with beam parameters as listed in Table~\ref{tab:beam_params_crossing_above}. PSB elements were modeled as thin lenses; thick elements were split using the Teapot algorithm into three slices, which gave satisfactory optics convergence. For the PIC space charge solver, tests showed that using more than $2\times10^5$ macroparticles results in an emittance growth smaller than $10^{-4}$~ mm mrad after tracking for $5\times10^4$ turns. Table~\ref{tab:simulation_params_crossing_above} summarizes the parameters used in the different simulation configurations.

\begin{table}[!htb]
    \caption{Typical simulation setup.}
    \centering
    \begin{tabular}{r l}
    \hline
    6D tracking code: & \texttt{Xsuite} \\
    Beam characteristics: & similar to Table~\ref{tab:beam_params_crossing_above} \\
    Number of macroparticles: & $10^6$ \\
    Magnetic element slices & $3$ \\
    Direct space charge solver: & PIC (2.5D) \\
    Space charge grid: & $128\times128\times64$ \\
    Space charge kicks per turn: & $160$ \\
    Tracked turns: & up to $3\times10^5$ \\
    \hline
    \end{tabular}
    \label{tab:simulation_params_crossing_above}
\end{table}

\section{Controlling the orientation of the beamlets}\label{appendix:orientation_of_beamlets}

A coasting beam with the parameters in Table~\ref{tab:beam_params_crossing_above}, was set to cross the half-integer resonance from above. At a fixed tune distance to the resonance, the vertical beam profile was measured for different resonance excitation settings.

\begin{figure}[!htb]
   \centering
   \includegraphics*[width=.6\columnwidth]{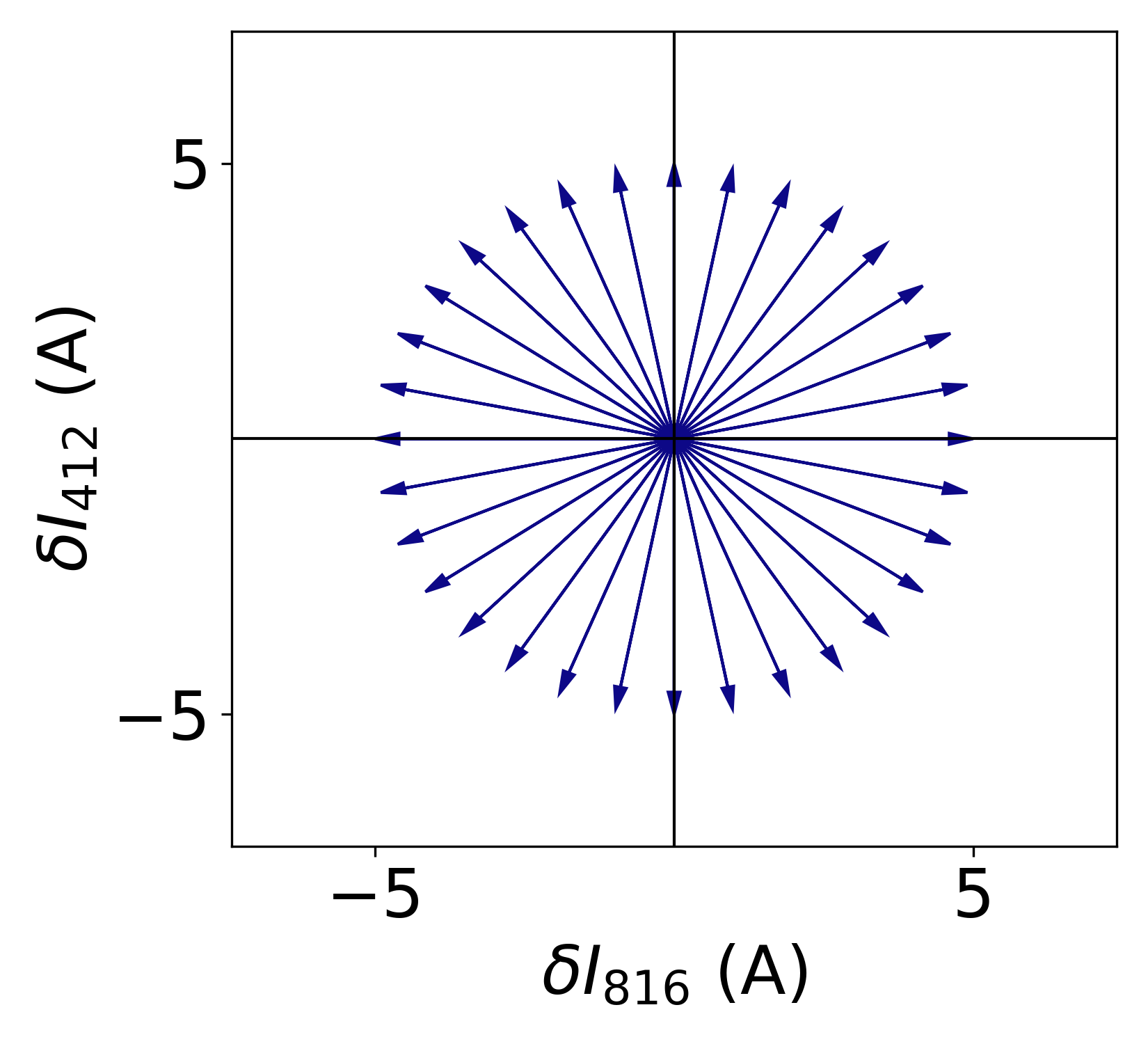}
   \includegraphics*[width=.7\columnwidth]{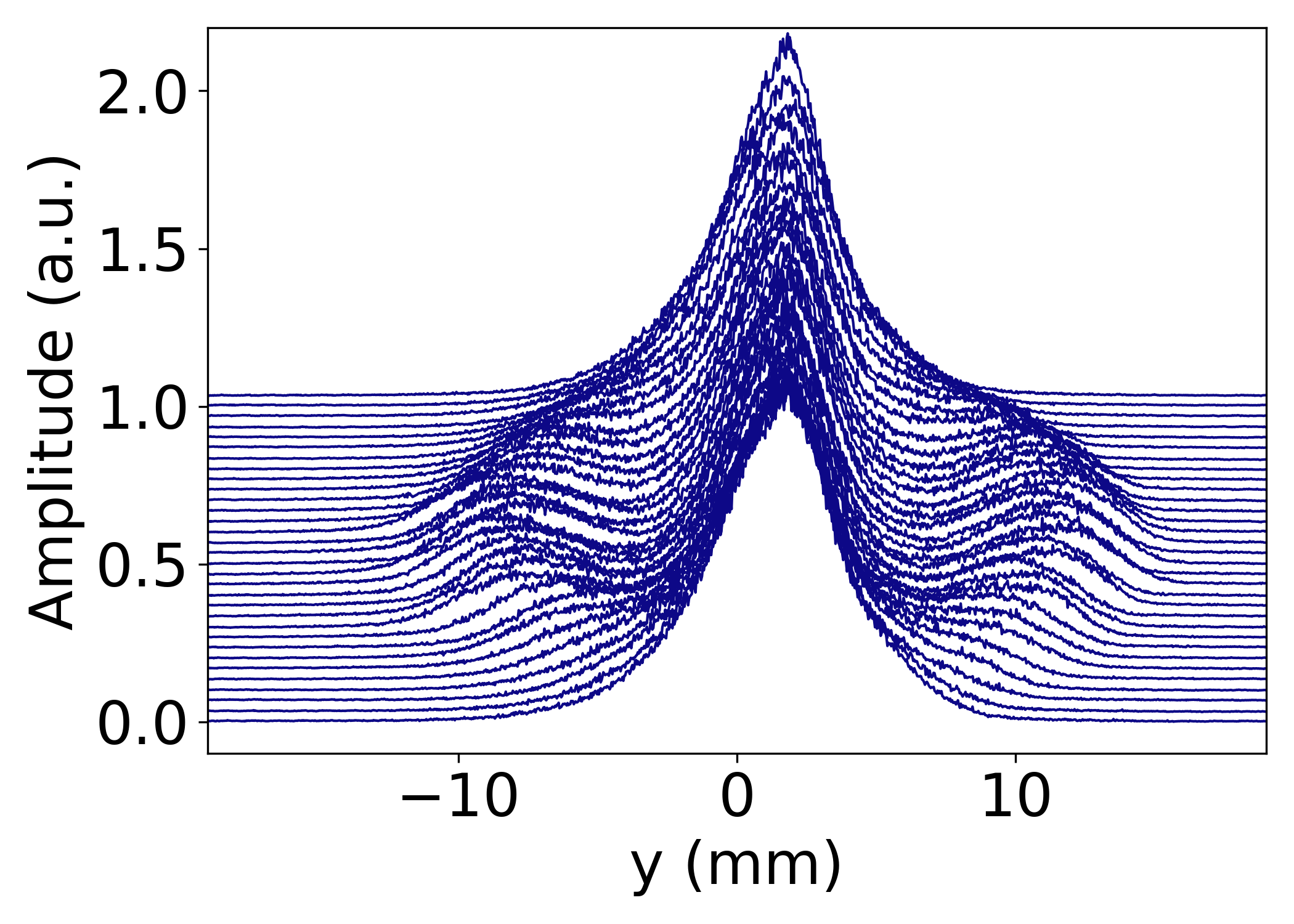}
   \caption{Top: illustration of the different excitation of the half-integer resonance (same amplitude but different phases). Bottom: measured vertical beam profiles at a fixed distance from the half-integer resonance for the different excitations (plotted as a mountainrange).}
   \label{fig:phase_space_rotation}
\end{figure}

The resonance was excited using both QNO816 and QNO412 quadrupole families of the PSB. These quadrupole corrector families are orthogonal in the resonance driving space of the half-integer resonance. The total excitation of the resonance was kept constant as:
\begin{equation}
    \delta I = \sqrt{\delta I_{816}^2 + \delta I_{412}^2},
\end{equation}
while varying the relative contribution of each family as shown in Fig.~\ref{fig:phase_space_rotation} (top). This configuration allowed the phase of the resonance driving term to be scanned over the full $0$–$2\pi$ range, while keeping its amplitude, and thus the resonance stopband width, constant.

The measured vertical profiles at a fixed distance from the resonance and for all phases are shown in Fig.~\ref{fig:phase_space_rotation} (bottom) as a mountain-range plot. Changing the relative strengths of the two families changes the position of the beamlets in the profiles. For certain excitations, the beamlets are not visible in the profiles and only the main Gaussian core is observed. This behavior can be understood by noticing that varying the relative strengths rotates the phase of the resonance driving term. This rotation changes the orientation of the phase-space islands and their trapped particles (as described in Sec.~\ref{subsec:particle_trapping}), thus modifying their projection in the vertical plane at the profile monitor. For certain phases, the beamlets overlap with the beam core and disappear from the measured profile.

For most of the studies presented in this work, the excitation setting $\delta I_{816}=-2$~A and $\delta I_{412}=0$~A was used, as it oriented the islands (i.e.~the beamlets) for maximum visibility at the wire scanner location.

\section{Remerging of trapped beamlets}\label{appendix:remerging}

\begin{figure}[htb]
   \centering
   \includegraphics*[width=.99\columnwidth]{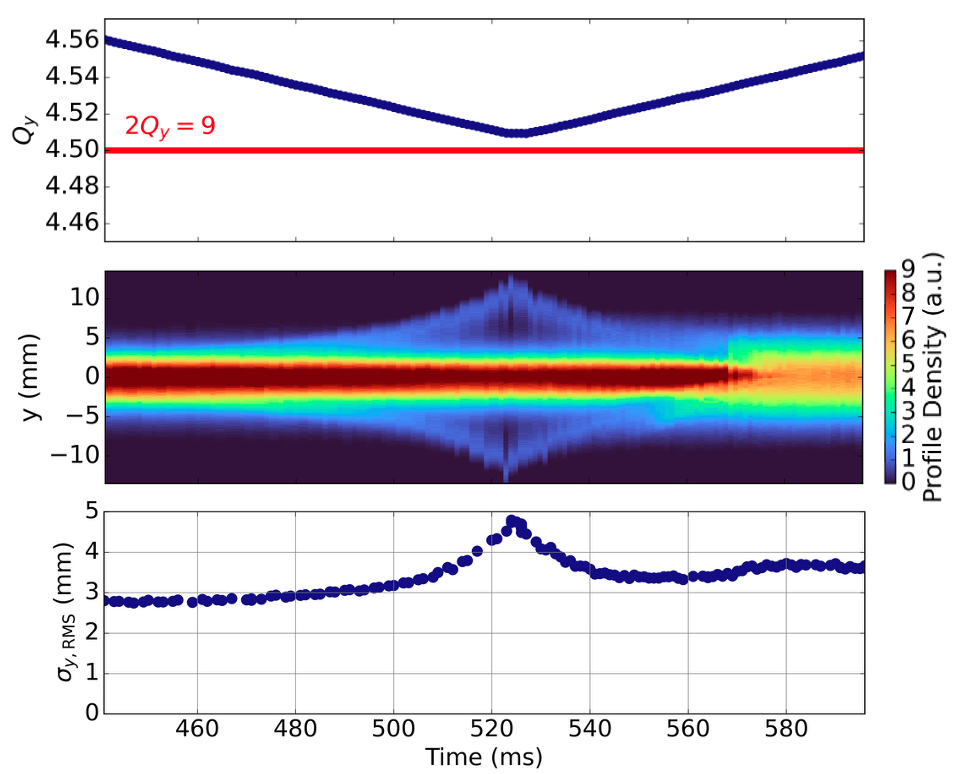}
   \caption{Top: measured vertical tune evolution. Middle: measured vertical beam profile evolution. Bottom: measured vertical beam size evolution.}
   \label{fig:remerging}
\end{figure}

An additional experiment was performed using the same beam parameters as in Table~\ref{tab:beam_params_crossing_above}. In this case, the vertical tune was ramped towards the resonance, as in Fig.~\ref{fig:tune_ramp}, but just before the beamlets reached the machine aperture, the tune ramp direction was reversed, moving the tune away from the resonance at the same rate as shown in Fig.\ref{fig:remerging} (top). During this process, vertical beam profiles were acquired every 1~ms ($\sim$1000 turns) as shown in Fig.\ref{fig:remerging} (middle). As the tune moved away from the resonance, the beamlets returned towards the beam core. However, once they reached the core edge, these particles filamented in phase space resulting in a larger beam profile compared to the initial beam profile. The final beam size thus remained slightly larger than the initial one, as shown in Fig.\ref{fig:remerging} (bottom).

\section{Limitations of the wire scanner measurements}
\label{appendix:wire-scattering}

The beam profiles were measured with a wire scanner~\cite{Veness:2289486}, a device that moves a thin wire across the beam over several turns. The beam-wire interaction generates a shower of secondary particles proportional to the density of beam particles, and detecting these secondary particles allows then to  reconstruct the beam profile. Two main effects can distort the measurements.

First, the finite time of the wire crossing the beam distribution, which spans several thousand turns at PSB injection energy. During relatively fast processes, such as the half-integer crossing from below as shown in Fig.~\ref{fig:crossing_below_profiles}, the beam distribution evolves considerably while the scan is in progress. Thus, the two sides of the beam profile correspond to different times in the evolution of the beam distribution. This effect is less pronounced when the beam distribution remains more static as in the case when the resonance crossing is slower, as shown in Fig.~\ref{fig:vertical_profile_single}.

Second, scattering of the beam particles from the beam–wire interaction can induce additional emittance blow-up during the scan, making one side of the profile appear broader, as observed in previous studies~\cite{ipac19-mopts101}.

\begin{figure}[htb]
   \centering
   \includegraphics*[width=.82\columnwidth]{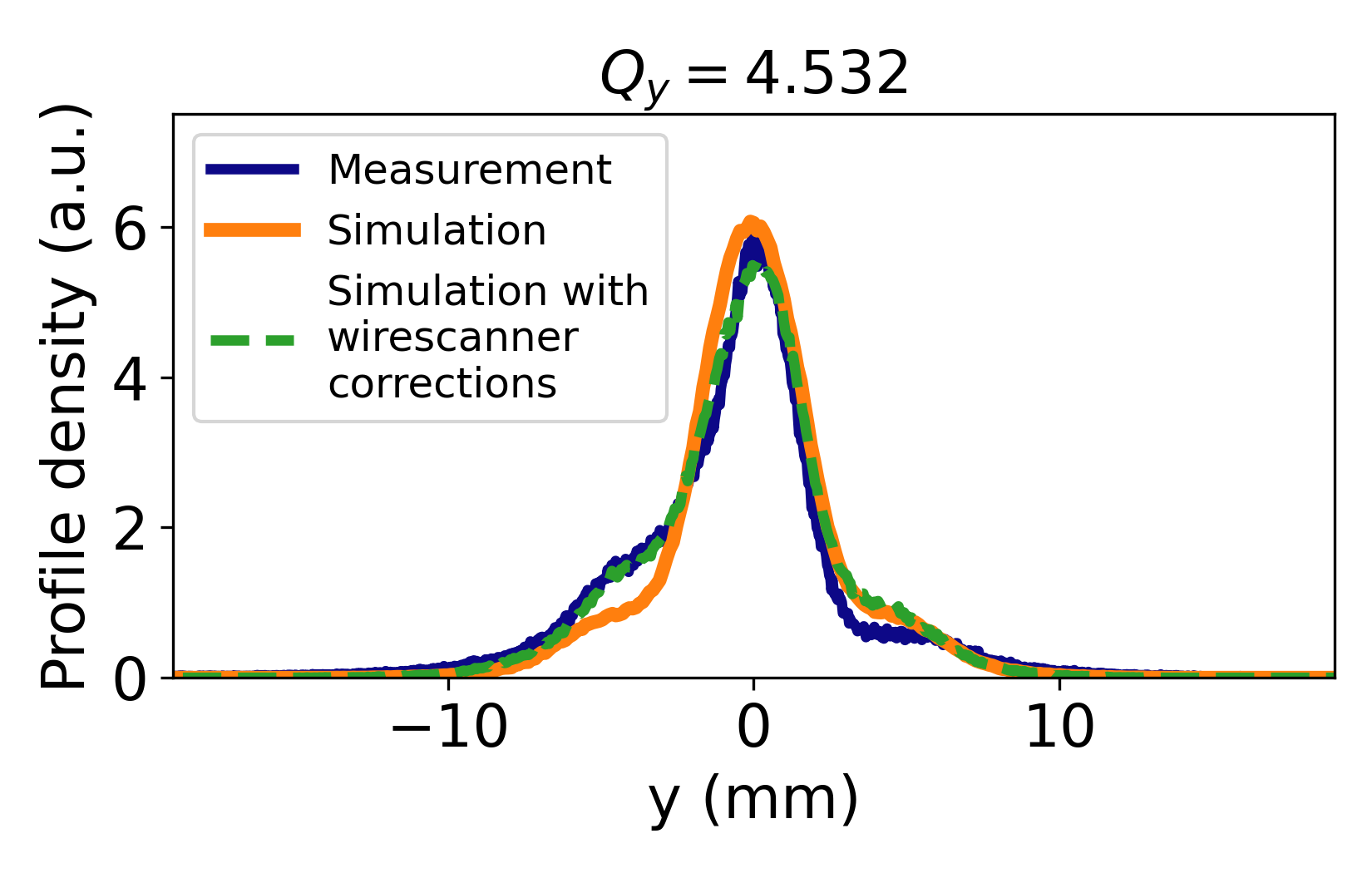}
   \caption{Measured vertical profile at $Q_y=4.532$ when crossing from below (blue, as in Fig.~\ref{fig:crossing_below_profiles}), simulated profile from phase space projection (orange), and simulated profile when including effects from the moving wire and the wire scattering (green).}
   \label{fig:wire_corrections}
\end{figure}

\begin{figure}[htb]
   \centering
   \includegraphics*[width=.82\columnwidth]{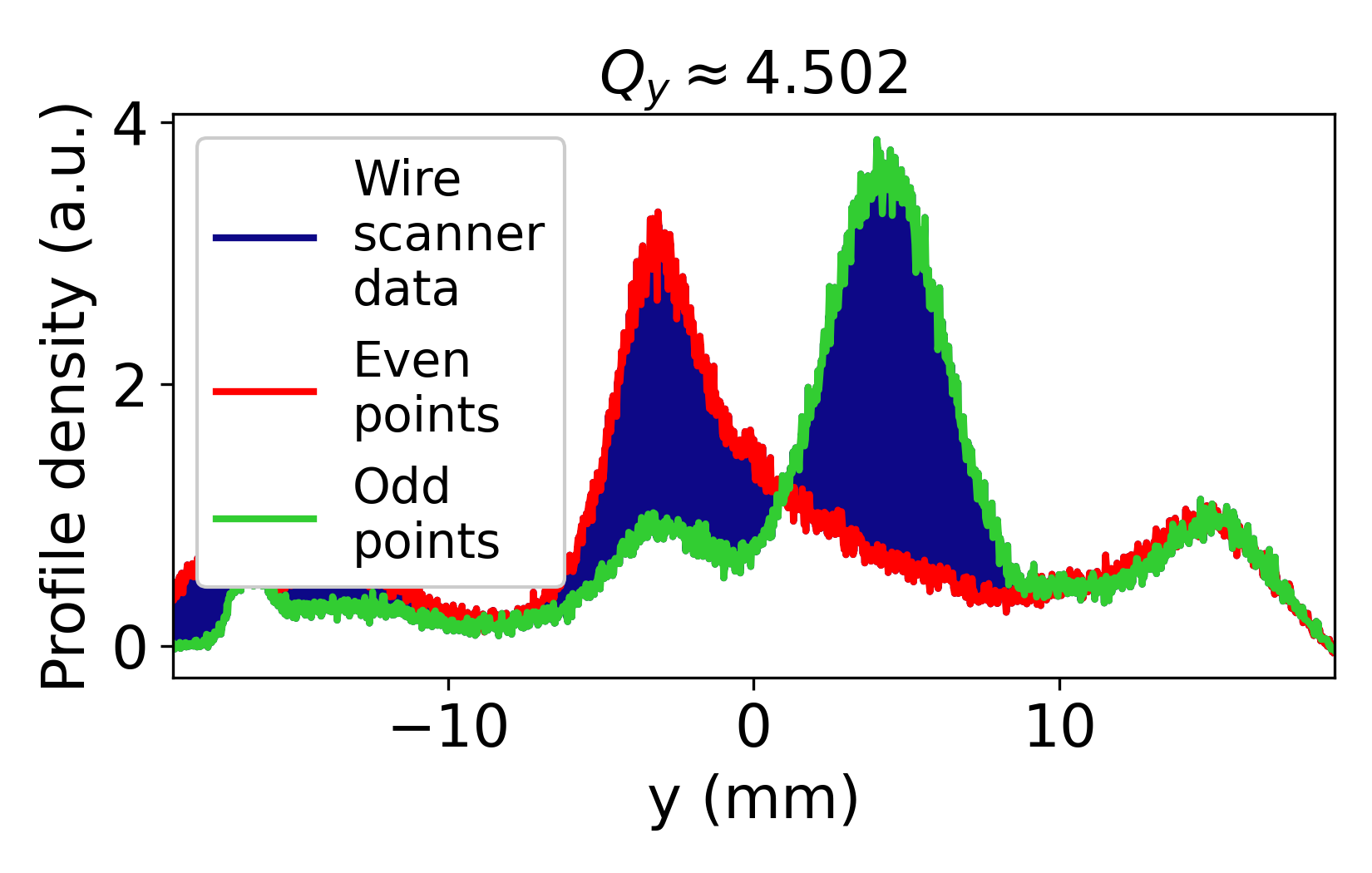}
   \caption{Measured vertical profile at $Q_y\approx 4.502$ when crossing from above (blue). The same data are shown separately for even (red) and odd (green) turns.}
   \label{fig:wire_centroid}
\end{figure}

Tracking simulations model the process of the wire scan including both the wire motion and the scattering of beam particles on the wire itself. The simulations reproduce profile asymmetries as those observed in the measured profiles very well, as shown in Fig.~\ref{fig:wire_corrections}. Thus, the asymmetry observed in the experimentally measured beam profiles is an artifact of the measurement device. The actual beam distribution without wire scan measurement is expected to be much more symmetric (as the orange profile in Fig.~\ref{fig:wire_centroid}). 

Interestingly, the finite time of the wire crossing the beam distribution can also provide complementary confirmation of the centroid oscillations. For instance, during the vertical crossing from above (Fig.~\ref{fig:vertical_profile_evolution}), the measured profile just before the complete beam loss ($Q_y \approx 4.502$) appears split into two peaks, as shown in Fig.~\ref{fig:wire_centroid}. Separating the data points from even and odd turns reveals that these peaks are consistent to the beam centroid performing betatron oscillations with a fractional tune of about $Q_y\approx0.5$ (Fig.~\ref{fig:intensity_tbt_measurement_simulation}). This provides an independent confirmation of the centroid oscillations observed with the turn-by-turn pickup.

\FloatBarrier

%\nocite{*}
%\bibliographystyle{apsrev4-2}
\bibliography{biblio}% Produces the bibliography via BibTeX.

\end{document}